\documentclass[preprint,nofootinbib,floats,eqsecnum,showpacs]{revtex4}

\usepackage{amsmath,amssymb,amsfonts}

\def\be{\begin{equation}}
\def\ee{\end{equation}}
\def\ba{\begin{eqnarray}}
\def\ea{\end{eqnarray}}
\def\nn{\nonumber}

\newcommand{\ket}[1]{\vert{#1} \rangle} 
\newcommand{\bra}[1]{\langle{#1}\,\vert} 
\newcommand{\oket}[1]{\vert{#1} )} 
\newcommand{\obra}[1]{({#1}\vert} 
\newcommand{\inner}[2]{{\langle {#1} \vert  {#2} \rangle}} 
\newcommand{\action}[2]{{( {#1}\vert  {#2} \rangle}} 
\newcommand{\oinner}[2]{{( {#1}\vert  {#2} )}} 
\newcommand{\opelem}[3]{{\langle {#1} \vert  {#2}  \vert  {#3} \rangle}} 

\newcommand{\secref}[1]{Sec.~\ref{#1}}

\newcommand{\eqnref}[1]{(\ref{#1})}

\newcommand{\appref}[1]{Appendix~\ref{#1}}
\newcommand{\footref}[1]{Footnote~\ref{#1}}

\begin{document}


\title{Timeless path integral for relativistic quantum mechanics}

\author{Dah-Wei Chiou}
\email{chiou@gravity.psu.edu}
\affiliation{${}^1$Department of Physics, Beijing Normal University, Beijing 100875, China\\
             ${}^2$Center for Condensed Matter Sciences, National Taiwan University, Taipei 10617, Taiwan}

\begin{abstract}
Starting from the canonical formalism of relativistic (timeless) quantum mechanics, the formulation of timeless path integral is rigorously derived. The transition amplitude is reformulated as the sum, or functional integral, over all possible paths in the constraint surface specified by the (relativistic) Hamiltonian constraint, and each path contributes with a phase identical to the classical action divided by $\hbar$. The timeless path integral manifests the timeless feature as it is completely independent of the parametrization for paths. For the special case that the Hamiltonian constraint is a quadratic polynomial in momenta, the transition amplitude admits the timeless Feynman's path integral over the (relativistic) configuration space. Meanwhile, the difference between relativistic quantum mechanics and conventional nonrelativistic (with time) quantum mechanics is elaborated on in light of timeless path integral.
\end{abstract}


\maketitle

\section{Introduction}
The idea that quantum mechanics can be well defined even if the notion of time is absent has been proposed \cite{Rovelli:1989jn,Rovelli:1988qp} and developed in a number of different strategies \cite{Rovelli:1990jm,Rovelli:1991ni,Reisenberger:2001pk,Marolf:2002ve,Hartle:1992as}. The motivation for formulating quantum mechanics in \emph{timeless} description comes from the research on quantum gravity, as in the quantum theory of general relativity, the spacetime background is not fixed and generally it is not possible to make sense of quantum variables ``at a moment of time''. This is closely related to the ``problem of time'' in quantum gravity~\cite{Isham:1992ms}.

In particular, a comprehensive formulation for the relativistic (timeless) quantum mechanics and its probabilistic interpretation are presented in Chapter 5 of \cite{Rovelli:book}.\footnote{The adjective ``relativistic'' connotes \emph{relational} correlations between physical variables in the timeless description. It should not be confused with the adjective for the theory of (special) relativity.} The formulation is based on the canonical (Hilbert spaces and self-adjoint operators) formalism and we wonder whether it also admits the covariant (sum-over-histories) formalism. In the conventional nonrelativistic (with time) quantum mechanics, the transition amplitudes are the matrix elements of the unitary evolution generated by the Hamiltonian and can be reformulated as the sum over histories, called the path integral. In the relativistic quantum mechanics, however, the concept of time evolution is not well defined in the fundamental level; therefore, conceptual issues and technical subtleties arise when one tries to derive the timeless path integral from the canonical formalism.
Various aspects of sum-over-histories approaches to relativistic quantum mechanics have been considered for a variety of models \cite{Hartle:1986yu,Hartle:1984ut,Brown:1989ne,Brown:1992bq,Teitelboim:1989fi,Gryb:2008rz, Halliwell:1992nj,Halliwell:2009rw,Halliwell:2012nj}.\footnote{For more references in the general area of ``timeless'' quantum theories, also see Refs 5 and 10-15 cited in \cite{Halliwell:2009rw} and Refs 11-21 cited in \cite{Halliwell:2012nj}.} Particularly, the method of path integral quantization elucidates the timelessness of the reparametrization-invariant quantum theory as the result of a superposition of clocks via Jacobi's principle \cite{Gryb:2008rz} and the existence of composition laws in relativistic quantum mechanics via path decomposition expansion \cite{Halliwell:1992nj}. However, rigorous derivation to the sum-over-histories formalism from a well-formulated canonical formalism is still lacking in relativistic quantum mechanics, and many important questions remain unclear such as what operator ordering has to be taken to yield a sensible path integral and what exactly the measure of the path integral is.

The aim of this paper is not to formulate a new sum-over-histories approach from a new perspective but instead to rigorously derive the timeless path integral for relativistic quantum mechanics, starting from the canonical formulation specifically described in \cite{Rovelli:book}. The main difficulty lies in the fact that the ordinary time-slicing process of path integrals cannot be directly carried over as there is no privileged observable to be treated as time, and thus new techniques have to be devised.\footnote{\label{foot:clarification}It should be noted that the path integral derived in this paper is fundamentally different from that in \cite{Hartle:1984ut}. While this paper gives the timeless path integral for \emph{relativistic} quantum mechanics, what in \cite{Hartle:1984ut} is the path integral for \emph{nonrelativistic} quantum mechanics recast into a parameterized theory, which manifests reparametrization invariance in an apparently timeless fashion. The underlying mechanics considered in \cite{Hartle:1984ut} is till the conventional (nonrelativistic) quantum mechanics, as opposed to the timeless (relativistic) quantum mechanics in this paper. The path integral in \cite{Hartle:1984ut}, although cast in a timeless fashion, is equivalent to the ordinary path integral of conventional quantum mechanics and thus no difficulty arises in regard to time-slicing. When the system is \emph{strictly deparametrizable}, the transition amplitude obtained from timeless path integral coincides with that from ordinary path integral, but their interpretations of probability and physics are profoundly different. The difference is subtle but crucial as discussed in \secref{sec:remarks on deparametrizable systems} and \secref{sec:deparametrizable systems}. In the same spirit of \cite{Hartle:1984ut}, it is suggested in \cite{Brown:1989ne} that, following the paradigm of the relationship between Jacobi's and Hamilton's action principles, one can correspond the Wheeler-DeWitt equation to a time-independent Schr\"{o}dinger equation and derive a time-dependent Wheeler-DeWitt equation of the Schr\"{o}dinger type with the four-volume of spacetime playing the role of physical time. (For this aspect of the problem of time, also see \cite{Gryb:2008rz} for the much simpler case of nonrelativistic particles moving through a space-dependent potential.) The resulting quantum theory of gravity (if can be constructed consistently) is again fundamentally different from that by directly quantizing the Wheeler-DeWitt equation in the manner of timeless quantum mechanics specifically formulated in \cite{Rovelli:book}. (Also see Appendix of \cite{Brown:1992bq} for the path integral for Jacobi's action based on the ideas of \cite{Brown:1989ne}.)} It turns out, nevertheless, the transition amplitude can be reformulated as the sum, or functional integral, over all possible paths in the constraint surface $\Sigma$ specified by the (relativistic) Hamiltonian constraint $H(q^a,p_a)=0$ for the configuration variables $q^a$ and their conjugate momenta $p_a$, and each path contributes with a phase identical to the classical action divided by $\hbar$. Unlike the conventional path integral in which every path is parameterized by the time variable $t$, the timeless path integral is completely independent of the parametrization for paths, manifesting the timeless feature. Furthermore, for the special case that the Hamiltonian constraint is a quadratic polynomial in $p_a$, the timeless path integral over $\Sigma$ reduces to the timeless Feynman's path integral over the (relativistic) configuration space.

The timeless path integral for relativistic quantum mechanics is appealing both conceptually and technically. Conceptually, timeless path integral offers an alternative interpretation of relativistic quantum fluctuations and is more intuitive than the canonical formalism for many aspects. It can give a new point of view about how the conventional quantum mechanics with time emerges within a certain approximation and thus may help to resolve the problem of time. Technically, timeless path integral provides tractable tools to compute (at least numerically or approximately) the transition amplitudes which otherwise remain formal in the canonical formalism, as various approximation methods in path integral approaches have been widely exploited and can be readily adapted to the timeless context.

In the research of loop quantum gravity (LQG), the sum-over-histories formulation is an active research area that goes under the name ``spin foam models'' (SFMs) (see \cite{Rovelli:book} and references therein for LQG and SFMs). In particular, over the past years, SFMs in relation to the kinematics of LQG have been clearly established \cite{Engle:2007uq,Freidel:2007py,Engle:2007wy,Kaminski:2009fm}. However, the Hamiltonian dynamics of LQG is far from fully understood, and although well motivated, SFMs have not been systematically derived from any well-established theories of canonical quantum gravity. Meanwhile, loop quantum cosmology (LQC) has recently been cast in a sum-over-histories formulation, providing strong support for the general paradigm underlying SFMs \cite{Ashtekar:2009dn,Ashtekar:2010ve}. In this paper, the timeless path integral is systematically derived from the canonical formalism of relativistic quantum mechanics, and we hope it will shed new light on the issues of the interplay between LQG/LQC and SFMs.

This paper is organized as follows. We begin with a review on the classical theory of relativistic mechanics in \secref{sec:classical theory} and then a review on the quantum theory of relativistic mechanics in \secref{sec:quantum theory}.\footnote{Readers who are familiar with the partial observable approach to timeless quantum mechanics as detailed in Chapters 3 and 5 of \cite{Rovelli:book} may skip \secref{sec:classical theory} and \secref{sec:quantum theory} and come back whenever necessary. To avoid any confusion with other timeless notions of quantum theories (recall \footref{foot:clarification}), it is advised to sill read \secref{sec:remarks on deparametrizable systems}.} The main topic is presented in \secref{sec:timeless path integral}, where the timeless path integral is derived and investigated in detail. Conclusions are summarized and discussed in \secref{sec:discussion}. Additionally, the stationary phase approximation for timeless path integral is included in \appref{app:stationary approximation} and, in order to compare with the timeless path integral, we re-derive the path integral for conventional quantum mechanics in \appref{app:path integral}.

\section{Classical theory of relativistic mechanics}\label{sec:classical theory}
The conventional formulation of classical mechanics treats the time $t$ on a special footing and therefore is not broad enough for general-relativistic systems, which treat time on the equal footing as other variables. To include general-relativistic systems, we need a a more general formulation with a new conceptual scheme. A timeless formulation for relativistic classical mechanics is proposed for this purpose and described in detail in Chapter 3 of \cite{Rovelli:book}, excerpts from which are presented in this section with some new materials added to give a review and define notations.

\subsection{Hamiltonian formalism}\label{sec:Hamiltonian formalism}
Let $\mathcal{C}$ be the \emph{relativistic configuration space} coordinatized by $q^a$ for $a=1,2,\cdots,d$ with $q^a$ being the \emph{partial observables} and $d$ being the dimension of $\mathcal{C}$. In nonrelativistic mechanics, one of the partial observables can be singled out and treated specially as the time $t$, i.e.\ $q^a=(t,q^i)$, but this separation is generally not possible for general-relativistic systems. An observation yields a complete set of $q^a$, which is called an \emph{event}. In nonrelativistic mechanics, an observation is a reading of the time $t$ together with other readings $q^i$.

Consider the cotangent space $\Omega=T^*\mathcal{C}$ coordinatized by $q^a$ and their momenta $p_a$. The space $\Omega$ carries a natural one-form $\tilde{\theta}=p_adq^a$. Once the kinematics (i.e.\ the space $\mathcal{C}$ of the partial observables $q^a$) is known, the dynamics is fully determined by giving a \emph{constraint surface} $\Sigma$ in the space $\Omega$. The constraint surface $\Sigma$ is specified by $H=0$ with a function $H:\Omega\rightarrow\mathbb{R}^k$. Denote $\tilde{\gamma}$ an unparameterized curve in $\Omega$ (observables and momenta) and $\gamma$ its projection to $\mathcal{C}$ (observables only). The physical motion is determined by the function $H$ via the following
\begin{quote}
\textbf{Variational principle.} A curve $\gamma$ in $\mathcal{C}$ is a \emph{physical motion} connecting the events $q_1^a$ and $q_2^a$, if $\tilde{\gamma}$ extremizes the action
\be\label{eqn:cl action}
S[\tilde{\gamma}]=\int_{\tilde{\gamma}}p_a\,dq^a
\ee
in the class of the curves $\tilde{\gamma}$ which satisfy
\be\label{eqn:cl constraint}
H(q^a,p_a)=0,
\ee
(i.e.\ $\tilde{\gamma}\in\Sigma$) and
whose projection $\gamma$ to $\mathcal{C}$ connect $q_1^a$ and $q_2^a$.
\end{quote}

If $k=1$, $H$ is a scalar function and called the \emph{Hamiltonian constraint}. If $k>1$, there is gauge invariance and $H$ is called the \emph{relativistic Hamiltonian}. The pair $(\mathcal{C},H)$ describes a relativistic dynamical system. All (relativistic and nonrelativistic) Hamiltonian systems can be formulated in this timeless formalism.

By parameterizing the curve $\tilde{\gamma}$ with a parameter $\tau$, the action \eqnref{eqn:cl action} reads as
\be\label{eqn:cl action with N}
S[q^a,p_a,N_i]=\int d\tau\left( p_a(\tau)\,\frac{d q^a(\tau)}{d\tau}
-N_i(\tau)H^i(q^a,p_a)\right),
\ee
where the constraint \eqnref{eqn:cl constraint} has been implemented with the Lagrange multipliers $N_i(\tau)$. Varying this action with respect to $N_i(\tau)$, $p_a(\tau)$ and $q^a(\tau)$ yields the constraint equation(s) \eqnref{eqn:cl constraint} together with the Hamilton equations:
\begin{subequations}\label{eqn:Hamilton eqs}
\ba
\frac{dq^a(\tau)}{d\tau}&=&N_j(\tau)\frac{\partial H^j(q^a,p_a)}{\partial p_a},
\\
\frac{dp_a(\tau)}{d\tau}&=&-N_j(\tau)\frac{\partial H^j(q^a,p_a)}{\partial q^a}.
\ea
\end{subequations}
For $k>1$, a motion is determined by a $k$-dimensional surfaces in $\mathcal{C}$ and different choices of the $k$ arbitrary functions $N_j(\tau)$ determine different curves and parametrizations on the single surface that defines a motion. For $k=1$, a motion is a 1-dimensional curve in $\mathcal{C}$ and different choices of $N(\tau)$ correspond to different parametrizations for the same curve. Different solutions of $q^a(\tau)$ and $p_a(\tau)$ for different choices of $N_j(\tau)$ are gauge-equivalent representations of the same motion and different choices of $N_j(\tau)$ have no physical significance.

Along the solution curve, the change rate of $H$ with respect to $\tau$ is given by
\ba
\frac{dH^i}{d\tau}
&=&\frac{dq^a}{d\tau}\frac{\partial H^i}{\partial q^a}
+\frac{dp_a}{d\tau}\frac{\partial H^i}{\partial p_a}
=N_j\frac{dH^j}{dp_a}\frac{\partial H^i}{\partial q^a}
-N_j\frac{dH^j}{dq^a}\frac{\partial H^i}{\partial p_a}\nn\\
&\equiv&N_j\,\lbrace H^i,H^j\rbrace.
\ea
To be consistent, the physical motion should remain on the constraint surface $\Sigma$. That is, $dH/d\tau$ has to vanish along the curve. Therefore, we must have the condition
\be\label{eqn:first class}
\left.\lbrace H^i,H^j\rbrace\right\vert_\Sigma=0,
\qquad\text{abbreviated as}\quad
\lbrace H^i,H^j\rbrace\approx0
\ee
for all $i$ and $j$. A function $F(q^a,p_a)$ defined in a neighborhood of $\Sigma$ is called \emph{weakly zero} if $\left.F\right\vert_\Sigma=0$ (abbreviated as $F\approx0$) and called \emph{strongly zero} if
\be
\left.F\right\vert_\Sigma=0
\quad\text{and}\quad
\left.\left(\frac{\partial F}{\partial q^a},
\frac{\partial F}{\partial p_a}\right)\right\vert_\Sigma=0,
\qquad\text{abbreviated as}\quad
F\simeq0.
\ee
It can be proven that $F\approx0$ implies $F\simeq f_iH^i$ for some functions $f_i(q^a,p_a)$. Consequently, we have
\be\label{eqn:first class 2}
\lbrace H^i,H^j\rbrace\simeq {f^{ij}}_k(q^a,p_a)\,H^k.
\ee
The condition \eqnref{eqn:first class} ensures all constraints $H^i$ to be \emph{first class}. (See \cite{Wipf:1993xg} for more about constrained systems and the concept of first class constraints.)

\subsection{Nonrelativistic mechanics as a special case}\label{sec:nonrelativistic mechanics}
The conventional nonrelativistic mechanics can also be formulated in the timeless framework as a special case. For the nonrelativistic systems, the relativistic configuration space has the structure $\mathcal{C}=\mathbb{R}\times\mathcal{C}_0$, where $\mathcal{C}_0$ is the conventional nonrelativistic configuration space; i.e., $q^a=(t,q^i)$ as one of the partial observables is identified as the time $t$. Correspondingly, the momenta read as $p_a=(p_t,p_i)$ with $p_t$ being the conjugate momentum of $t$ and $p_i$ being the conjugate momenta of $q^i$. The Hamiltonian constraint is given by
\be\label{eqn:nonrel H}
H(t,q^i,p_t,p_i)=p_t+H_0(q^i,p_i;t),
\ee
where $H_0(q^i,p_i;t)$ is the conventional nonrelativistic Hamiltonian function. Given the Hamiltonian constraint in the form of \eqnref{eqn:nonrel H}, the Hamilton equations \eqnref{eqn:Hamilton eqs} lead to
\begin{subequations}
\ba
\frac{dt}{d\tau}=N(\tau),
\qquad
\frac{dp_t}{d\tau}=-N(\tau)\frac{\partial H_0}{\partial t},
\\
\frac{dq^i}{d\tau}=N(\tau)\frac{\partial H_0}{\partial p_i},
\qquad
\frac{dp_i}{d\tau}=-N(\tau)\frac{\partial H_0}{\partial q^i},
\ea
\end{subequations}
which read as
\be
\frac{dp_t}{dt}=-\frac{\partial H_0}{\partial t}
\ee
and
\be\label{eqn:nonrel Hamilton eqs}
\frac{dq^i}{dt}=\frac{\partial H_0}{\partial p_i}, \qquad
\frac{dp_i}{dt}=-\frac{\partial H_0}{\partial q^i},
\ee
if particularly we use $t$ to parameterize the curve of solutions.
Furthermore, the constraint \eqnref{eqn:cl constraint} dictates $p_t=-H_0$. Thus, the momentum $p_t$ is the negative of energy and it is a constant of motion if $H_0=H_0(q^i,p_i)$ has no explicit dependence on $t$. The equations in \eqnref{eqn:nonrel Hamilton eqs} are precisely the conventional Hamilton equations for nonrelativistic mechanics.

The Hamilton equations in \eqnref{eqn:nonrel Hamilton eqs} form a system of first-order ordinary differential equations. Given the initial condition $q^i(t_0)=q_0^i$ and $p_i(t_0)=p_{i0}$ at the time $t_0$, the existence and uniqueness theorem for ordinary differential equations states that there exists a solution of \eqnref{eqn:nonrel Hamilton eqs} given by $q^i=q^i(t)$ and $p_i=p_i(t)$ for $t\in\mathbb{R}$, and furthermore the solution is unique.\footnote{In order to apply the existence and uniqueness theorem, we assume $\partial H_0/\partial q^i$, $\partial H_0/\partial p_i$, $\partial^2 H_0/\partial {q^i}^2$, $\partial^2 H_0/\partial {p_i}^2$ and $\partial^2 H_0/\partial q^i \partial p_j$ all continuous.} As a consequence, $q^i$ and $p_i$ evolve as functions of $t$, and a physical motion is an \emph{open} curve in $\mathcal{C}=\mathbb{R}\times\mathcal{C}_0$, along which the observable $t$ is \emph{monotonic}.

A dynamical system in which a particular partial observable can be singled out as $t$ such that the Hamiltonian is separated as in the form of \eqnref{eqn:nonrel H} is called \emph{deparametrizable}. For deparametrizable systems, the change of $t$ is in accord with the ordinary notion of time, which does not turn around but grows monotonically along the physical motion. Generically, however, relativistic systems might be non-deparametrizable --- no preferred observable can serve as the time such that other variables are described as functions of time along the physical motion. The classical theory predicts the physical motion as an unparameterized curve, which gives \emph{correlations} between physical variables, not the way physical variables evolve with respect to a preferred time variable. In the next subsection, we will introduce the timeless double pendulum as an example to illustrate the timeless feature.

\subsection{Example: Timeless double pendulum}\label{sec:timeless double pendulum}
Let us now introduce a genuinely timeless system as a simple model to illustrate the mechanics without time. This model was first introduced in \cite{Rovelli:1990jm,Rovelli:1991ni} and used repeatedly as an example in \cite{Rovelli:book}.

Consider a mechanical system with two partial observables, $a$ and $b$, whose dynamics is specified by the relativistic Hamiltonian
\be
H(a,b,p_a,p_b)=\frac{1}{2}\left(p_a^2+p_b^2+a^2+b^2-2E\right)
\ee
with a given constant $E$. The relativistic configuration space is $\mathcal{C}=\mathbb{R}^2$ coordinatized by $a$ and $b$, and the cotangent space $\Omega=T^*\mathcal{C}$ is coordinatized by $(a,b,p_a,p_b)$. The constraint surface $\Sigma$ is specified by $H=0$; it is a 3-dimensional sphere of radius $\sqrt{2E}$ in $\Omega$.

In the $N(\tau)=1$ gauge, the Hamilton equations \eqnref{eqn:Hamilton eqs} give
\be
\frac{da}{d\tau}=p_a,
\qquad
\frac{db}{d\tau}=p_b,
\qquad
\frac{dp_a}{d\tau}=-a,
\qquad
\frac{dp_b}{d\tau}=-b,
\ee
and the Hamiltonian constraint \eqnref{eqn:cl constraint} gives
\be
a^2+b^2+p_a^2+p_b^2=2E.
\ee
The general solution is given by
\be
a(\tau)=A_a\sin(\tau), \qquad b(\tau)=A_b\sin(\tau+\beta),
\ee
where $A_a=\sqrt{2E}\sin\alpha$ and $A_b=\sqrt{2E}\cos\alpha$, and $\alpha$ and $\beta$ are constants.

Therefore, physical motions are closed curves (ellipses) in $\mathcal{C}=\mathbb{R}^2$. (Choosing different gauges for $N$ yields the same curve with different parametrizations.) This system is non-deparametrizable and does not admit a conventional Hamiltonian formulation, because, as discussed in \secref{sec:nonrelativistic mechanics}, physical motions in $\mathcal{C}=\mathbb{R}\times\mathcal{C}_0$ for a nonrelativistic system are monotonic in $t$ and thus cannot be closed curves.

\subsection{Lagrangian formalism}
Consider the special case that the relativistic Hamiltonian is given in the form:\footnote{In this subsection, the repeated index $a$ is not summed unless $\sum_a$ is explicitly used.}
\be\label{eqn:H special form}
H(q^a,p_a)=\sum_a\alpha_ap_a^2 + \sum_a\beta_ap_aq^a + \sum_a\gamma_ap_a +V(q^a),
\ee
where $\alpha_a$, $\beta_a$ and $\gamma_a$ are constant coefficients, and $V(q^a)$ is the potential which depends only on $q^a$. This form is quite generic and many examples of interest belong to this category such as the relativistic particle (free or subject to an external potential), the timeless double pendulum (harmonic or anharmonic) and the nonrelativistic system as described by \eqnref{eqn:nonrel H} with $H_0=\sum_ip_i^2/2m_i+V(q^i,t)$. The Hamilton equations \eqnref{eqn:Hamilton eqs} yields
\begin{subequations}\label{eqn:Hamilton eqs special case}
\ba
\label{eqn:Hamilton eqs special case a}
\frac{dq^a}{d\tau}&=&N\left(2\alpha_ap_a+\beta_aq^a+\gamma_a\right),\\
\frac{\;dp_a}{d\tau}&=&-N\left(\beta_ap_a+\frac{\partial V}{\partial q^a}\right).
\ea
\end{subequations}

Equation \eqnref{eqn:Hamilton eqs special case a} gives the relation between the momenta $p_a$ and the ``velocities'' $\dot{q}^a:={dq^a}/{d\tau}$, through which
the inverse Legendre transform recasts the action \eqnref{eqn:cl action with N} in terms of the Lagrangian function:
\ba\label{eqn:cl Lagrangian}
&&S[q^a,\dot{q}^a,N;\tau]=
\int d\tau\, L(q^a,\dot{q}^a,N)\nn\\
&=&\int d\tau
\left(\sum_a\frac{N}{4\alpha_a}\left[\frac{\dot{q}^a}{N}-\beta_aq^a-\gamma_a\right]^2
-NV(q^a)\right).
\ea
Variation with respect to $N$ yields
\ba\label{eqn:var to N on L}
\frac{\delta S}{\delta N}&\equiv&\frac{\partial L}{\partial N}=0
\quad\Rightarrow\nn\\
0&=&\sum_a\frac{1}{4\alpha_a}\left[\frac{\dot{q}^a}{N}-\beta_aq^a-\gamma_a\right]^2
-\sum_a\frac{\dot{q}^a}{2\alpha_aN} \left[\frac{\dot{q}^a}{N}-\beta_aq^a-\gamma_a\right]
-V\nn\\
&=&-\left(\sum_a\alpha_ap_a^2 + \sum_a\beta_ap_aq^a + \sum_a\gamma_ap_a +V\right)
=-H,
\ea
which is precisely the Hamiltonian constraint \eqnref{eqn:cl constraint}. On the other hand, variation with respect to $q^a$ gives the equation of motion as a second-order differential equation:
\ba\label{eqn:var to q on L}
&&\frac{\delta S}{\delta q^a}
\equiv\frac{\partial L}{\partial q^a}
-\frac{d}{d\tau}\frac{\partial L}{\partial\dot{q}^a}=0\nn\\
&\Rightarrow&\quad
\frac{d}{Nd\tau}\left(\frac{dq^a}{Nd\tau}\right)
=\beta_a^2q^a+\beta_a\gamma_a-2\alpha_a\frac{\partial V}{\partial q^a},
\ea
which is equivalent to \eqnref{eqn:Hamilton eqs special case}.

\section{Quantum theory of relativistic mechanics}\label{sec:quantum theory}
The timeless formulation for relativistic classical mechanics is reviewed in \secref{sec:classical theory}. Based on the Hamiltonian framework of the classical theory, the quantum theory of relativistic mechanics can be formulated in canonical formalism. Unlike the conventional quantum theory, relativistic quantum mechanics does not describe evolution in time, but correlations between observables.

In \secref{sec:general scheme}, we stipulate a general scheme for relativistic quantum mechanics, which is excerpted from Chapter 5 of \cite{Rovelli:book}. In \secref{sec:remarks on deparametrizable systems}, we comment on the difference between relativistic quantum mechanics and conventional quantum mechanics when the system is deparametrizable. In \secref{sec:timeless double pendulum qm}, as excerpted from Chapter 5 of \cite{Rovelli:book} again, we take the timeless double pendulum as a simple example to illuminate the timeless formalism. Issues on the physical Hilbert space are detailed in \secref{sec:physical Hilbert space} and the physical interpretations of quantum measurements and collapse are discussed in \secref{sec:measurements and collapse}.

\subsection{General scheme}\label{sec:general scheme}
Let $\mathcal{C}$ be the relativistic configuration space for the classical theory as described in the \secref{sec:Hamiltonian formalism}. The corresponding quantum theory can be formulated timelessly in the following scheme:
\begin{description}

\item[Kinematical states.] Let $\mathcal{S}\subset\mathcal{K}\subset\mathcal{S}'$ be the Gelfand triple defined over $\mathcal{C}$ with the measure $d^dq_a\equiv dq^1dq^2\cdots dq^d$.\footnote{That is, $\mathcal{S}$ is the space of the smooth functions $f(q^a)$ on $\mathcal{C}$ with fast decrease, $\mathcal{K}=L^2[\mathcal{C},d^dq^a]$ is a Hilbert space, and $\mathcal{S}'$ is formed by the tempered distributions over $\mathcal{C}$.} The kinematical states of a system are represented by vectors $\ket{\psi}\in\mathcal{K}$, and $\mathcal{K}$ is called the \emph{kinematical Hilbert space}.

\item[Partial observables.] A partial observable is represented by a self-adjoint operator in $\mathcal{K}$. The simultaneous eigenstates $\ket{s}$ of a complete set of commuting partial observables are called \emph{quantum events}. In particular, $\hat{q}^a$ and $\hat{p}_a$ are partial observables acting respectively as multiplicative and differential operators on $\psi(q^a)$; i.e., $\hat{q}^a\psi(q^a)=q^a\psi(q^a)$ and $\hat{p}_a\psi(q^a)=-i\hbar\,\partial\psi(q^a)/\partial q^a$. Their eigenstates $\ket{q^a}$ (defined as $\hat{q}^a\ket{q^a}=q^a\ket{q^a}$) and $\ket{p_a}$ (defined as $\hat{p}_a\ket{p_a}=p_a\ket{p_a}$) are both quantum events.

\item[Dynamics.] Dynamics is defined by a self-adjoint operator $\hat{H}$ in $\mathcal{K}$, called \emph{relativistic Hamiltonian}. The operator from $\mathcal{S}$ to $\mathcal{S}'$ schematically defined as
    \be\label{eqn:projector}
    \hat{P}=\int d\tau\, e^{-i\tau\hat{H}}
    \ee
    is called the ``projector''.\footnote{The integration range depends on the system: It is over a compact space if the spectrum of $\hat{H}$ is discrete and over a noncompact space if the spectrum is continuous. The operator $\hat{P}$ is a projector in the precise sense only if zero is a part of the discrete spectrum of $\hat{H}$.} The matrix elements
    \be\label{eqn:W s s'}
    W(s,s'):=\opelem{s}{\hat{P}}{s'}
    \ee
    are called \emph{transition amplitudes}, which encode entire physics of the dynamics.

\item[Physical states.] A \emph{physical state} is a solution of the quantum Hamiltonian constraint equation:
    \be\label{eqn:qm constraint}
    \hat{H}\ket{\psi}=0,
    \ee
    which is the quantum counterpart of \eqnref{eqn:cl constraint}. Given an arbitrary kinematical state $\ket{\psi_\alpha}\in\mathcal{S}$, we can associate an element $\obra{\Psi_{\psi_\alpha}}\in\mathcal{S}'$, defined by its (linear) action on arbitrary states $\ket{\psi_\beta}\in\mathcal{S}$ as
    \be
    \action{\Psi_{\psi_\alpha}}{\psi_\beta}=\int d\tau\, \inner{e^{i\tau\hat{H}}\psi_\alpha}{\psi_\beta}
    \equiv {\opelem{\psi_\alpha}{\hat{P}}{\psi_\beta}},
    \ee
    such that $\obra{\Psi_{\psi_\alpha}}$ is a physical state, namely, a solution to \eqnref{eqn:qm constraint}. The solution space is endowed with the Hermitian inner product:
    \be\label{eqn:physical inner product}
    \oinner{\Psi_{\psi_\alpha}}{\Psi_{\psi_\beta}}
    :=\action{\Psi_{\psi_\alpha}}{\psi_\beta},
    \ee
    which is called the \emph{physical inner product}. The Cauchy completion of the solution space with respect to the physical inner product $\oinner{\cdot}{\cdot}$ is called the \emph{physical Hilbert space} and denoted as $\mathcal{H}$.

\item[Measurements and collapse.] If the measurement corresponding to a partial observable $\hat{A}$ is performed, the outcome takes the value of one of the eigenvalues of $\hat{A}$ if the spectrum of $\hat{A}$ is discrete, or in a small spectral region (with uncertainty) if the spectrum is continuous. Measuring a complete set of partial observables $\hat{A}_i$ \emph{simultaneously} is called a \emph{complete measurement} at an ``instance'',\footnote{In the timeless language, a complete measurement is said to be conducted at some ``instance'', not at some ``instant''.} the outcome of which gives rise to a kinematical state $\ket{\psi_\alpha}$ (which is a simultaneous eigenstate of $\hat{A}_i$ if the spectra of $\hat{A}_i$ are discrete). The physical state is said to be \emph{collapsed} to $\oket{\Psi_{\psi_\alpha}}$ by the complete measurement.

\item[Prediction in terms of probability.] If at one instance a complete measurement yields $\ket{\psi_\alpha}$, the probability that at another instance another complete measurement yields $\ket{\psi_\beta}$ is given by
    \be
    \mathcal{P}_{\beta\alpha}=
    \left\vert
    \frac{\oinner{\Psi_{\psi_\beta}}{\Psi_{\psi_\alpha}}}
    {\sqrt{\oinner{\Psi_{\psi_\beta}}{\Psi_{\psi_\beta}}}\
    \sqrt{\rule{0mm}{3.6mm}\oinner{\Psi_{\psi_\alpha}}{\Psi_{\psi_\alpha}}}}
    \right\vert^2
    =\left\vert
    \frac{W[\psi_\beta,\psi_\alpha]}
    {\sqrt{\rule{0mm}{3.5mm}W[\psi_\beta,\psi_\beta]}\ \sqrt{W[\psi_\alpha,\psi_\alpha]}}
    \right\vert^2,
    \ee
    where
    \be
    W[\psi_\beta,\psi_\alpha]:=\opelem{\psi_\beta}{\hat{P}}{\psi_\alpha}
    =\int ds \int ds'\ \overline{\psi_\beta(s)}\ W(s,s') \, \psi_\alpha(s').
    \ee
    In particular, if the quantum events $s$ make up a discrete spectrum, the probability of the quantum event $s$ given the quantum event $s'$ is
    \be\label{eqn:P s s'}
    \mathcal{P}_{ss'}
    =\left\vert
    \frac{W(s,s')}{\sqrt{W(s,s)}\ \sqrt{W(s',s')}}
    \right\vert^2.
    \ee
    If the spectrum is continuous, the probability of a quantum event in a small spectral region $R$ given a quantum event in a small spectral region $R'$ is
    \be\label{eqn:P R R'}
    \mathcal{P}_{RR'}
    =\left\vert
    \frac{W(R,R')}{\sqrt{W(R,R)}\ \sqrt{W(R',R')}}
    \right\vert^2,
    \ee
    where
    \be\label{eqn:W R R'}
    W(R,R'):=\int_R ds \int_{R'} ds'\ W(s,s').
    \ee

\end{description}

The general scheme stipulated above gives a sound (axiomatic) framework for relativistic quantum mechanics. However, it is far from complete and remains provisional as many aspects need to be further clarified. One obvious problem is what precisely the small region $R$ in \eqnref{eqn:W R R'} should be associated to when a complete measurement is conducted in the case of a continuous spectrum. Additionally, although simultaneous \emph{exact} measurements of non-commuting (partial) observables are impossible, may simultaneous \emph{inaccurate} measurements of them (known as ``joint measurements'') still be possible? If yes, given outcomes with inaccuracies, what exact  kinematical state $\ket{\psi_\alpha}$ does the joint measurement yield? Furthermore, what are correct treatments for the measurements which are performed not at an instance but over a short continuous duration or repeatedly performed at successive instances?\footnote{These problems already exist in the orthodox formulation of conventional quantum mechanics. A lot studies of research have been devoted to these issues and vigorous debates remain unsettled. Among them, see \cite{Arthurs-Kelly} for the simultaneous measurement of a pair of conjugate observables and \cite{Uffink} for the ``joint measurement problem''. We hope these issues would gain more insight in light of relativistic quantum mechanics.}

\subsection{Remarks on deparametrizable systems}\label{sec:remarks on deparametrizable systems}
It should be emphasized that, unlike the classical theory, the relativistic quantum mechanics formulated in \secref{sec:general scheme} is \emph{not} equivalent to the conventional quantum theory, even if the system is deparametrizable. In conventional quantum mechanics, the time $t$ is treated as a parameter and not quantized as an operator. Thus, the measurement of $t$ is presumed to have zero uncertainty ($\Delta t=0$). In relativistic quantum mechanics, by contrast, $t$ is on the same footing as other observables $q^i$ and the measurement of $t$ will yield nonzero $\Delta t$.

If a system is deparametrizable and particularly $H_0$ in \eqnref{eqn:nonrel H} is not explicitly dependent on $t$, we have $\hat{H}=\hat{p}_t+\hat{H}_0(\hat{q}^i,\hat{p}_i)$ and the projector $\hat{P}$ can be cast as
\ba
\hat{P}&:=&\int d\tau\, e^{-i\tau\hat{H}}
=\int d\tau dp_t\, dE\, e^{-i\tau\hat{H}} \ket{p_t,E}\bra{p_t,E}\nonumber\\
&=&\int d\tau dp_t\, dE\, e^{-i\tau(p_t+E)} \ket{p_t,E}\bra{p_t,E}\nonumber\\
&\propto&\int dp_t\, dE\, \delta(p_t+E) \ket{p_t,E}\bra{p_t,E}
=\int dE\,\ket{-E,E}\bra{-E,E},
\ea
where $\ket{p_t,E}\equiv\ket{p_t}\otimes\ket{E}$ are simultaneous eigenstates of $\hat{p}_t$ and $\hat{H}_0$ with eigenvalues $p_t$ and $E$ (note that $\hat{p}_t$ and $\hat{H}_0$ commute).
Consequently, the transition amplitude for relativistic quantum mechanics is given by
\ba\label{eqn:W strict}
&&W(q^a,q'^a)\equiv W(t,q^i,t',q'^i):=\opelem{q^a}{\hat{P}}{q'^a}\nonumber\\
&\propto& \int dE \, \inner{t,q^i}{-E,E}\inner{-E,E}{t',q'^i}
= \int dE \, e^{-iE(t-t')}f_E(q^i)\overline{f_E(q'^i)}\,,
\ea
where $f_E(q^i):=\inner{q^i}{E}$ is the eigenfunction of $\hat{H}_0$ in $q^i$-representation.
On the other hand, the transition amplitude for conventional quantum mechanics is given by
\ba
&&G(q^i,t;q'^i,t'):=\opelem{q^i}{e^{-i\hat{H}_0(t-t')}}{q'^i}\nonumber\\
&=&\int dE\, dE'\, \inner{q^i}{E}\opelem{E}{e^{-i\hat{H}_0(t-t')}}{E'}\inner{E'}{q'^i}
= \int dE \, e^{-iE(t-t')}f_E(q^i)\overline{f_E(q'^i)}\,,
\ea
which happens to be identical to $W(t,q^i,t',q'^i)$ in \eqnref{eqn:W strict}.
In fact, even if $H_0(q^i,p_i;t)$ depends on $t$ explicitly, as long as $[\hat{H}_0(t_1),\hat{H}_0(t_2)]=0$ for all $t_1, t_2$, $W(t,q^i,t',q'^i)$ and $G(t,q^i;t',q'^i)$ are identical to each other (up to an irrelevant normalization factor for $W$) as will be proven in light of path integral in \secref{sec:deparametrizable systems}. A system is said to be \emph{strictly deparametrizable} if $[\hat{H}_0(t_1),\hat{H}_0(t_2)]=0$.\footnote{For deparametrizable systems, $[\hat{H}_0(\hat{q}^i,\hat{p}_i,\hat{t}_1),\hat{H}_0(\hat{q}^i,\hat{p}_i,\hat{t}_2)]=0$ if and only if $[\hat{H}_0(\hat{q}^i,\hat{p}_i,t_1),\hat{H}_0(\hat{q}^i,\hat{p}_i,t_2)]=0$.}
For strictly deparametrizable systems, in a sense, relativistic quantum mechanics and conventional quantum mechanics are different at the level of kinematics (observation and measurement) but identical at the level of dynamics (transition amplitudes).

In conventional quantum mechanics, given a (non-relativistic) quantum event in a small spectral $R'_0$ measured at time $t'$ with small time inaccuracy $\Delta t'$, the (averaged) probability of the quantum event in a small spectral $R_0$ measured at time $t$ with small time inaccuracy $\Delta t$ can be prescribed as
\ba\label{eqn:P R0 R0'}
&&\mathcal{P}_{R_0,t\pm\Delta t;R'_0,t'\pm\Delta t'}\nn\\
&=&\frac{1}{\Delta t\,\Delta t'}\int_{t-\Delta t}^{t+\Delta t}dt\int_{t'-\Delta t'}^{t'+\Delta t'}dt'
\left|\int_{R_0}d^{\,d-1}q^i \int_{R'_0}d^{d-1}q'^i\, G(q^i,t;q'^i,t')\right|^2,
\ea
which is different from \eqnref{eqn:P R R'} with $R=[t-\Delta t,t+\Delta t]\times R_0$ and $R'=[t'-\Delta t',t'+\Delta t']\times R'_0$ even though $W(t,q^i,t',q'^i)\propto G(q^i,t;q'^i,t')$, as amplitudes are summed over time \emph{interferentially} in \eqnref{eqn:P R R'} while they are summed over time \emph{additively} in \eqnref{eqn:P R0 R0'}. Only if $\Delta t$ and $\Delta t'$ are small enough, the interference in time can be neglected.
Particularly, for a simple harmonic oscillator govern by the relativistic Hamiltonian $H=p_t+H_0=p_t+{p_\alpha^2}/{2m}+{m\omega^2\alpha^2}/{2}$, it was shown in \cite{Reisenberger:2001pk,Marolf:2002ve} that,  if $\Delta t\ll m\Delta\alpha^2/\hbar$, we can ignore the temporal resolution $\Delta t$ and idealize the measurement of $t$ as \emph{instantaneous}, and the conventional nonrelativistic quantum theory is recovered as a good approximation of the relativistic quantum mechanics.

On the other hand, for \emph{non-strictly} deparametrizable systems, i.e.\ $[\hat{H}_0(t_1),\hat{H}_0(t_2)]\neq0$, relativistic quantum mechanics and conventional quantum mechanics are different both for kinematics and dynamics. For many situations of interest, nevertheless, $G(t,q^i;t',q'^i)$ can be regarded as a reasonable approximation of $W(t,q^i,t',q'^i)$ as will be discussed in \secref{sec:deparametrizable systems}.

\subsection{Example: Timeless double pendulum}\label{sec:timeless double pendulum qm}
Take the timeless double pendulum introduced in \secref{sec:timeless double pendulum} as an example. The kinematical Hilbert space is $\mathcal{K}=L^2(\mathbb{R}^2,dadb)$, and the quantum Hamiltonian equation reads as
\be\label{eqn:qm constraint for double pendulum}
\hat{H}\psi(a,b)=
\frac{1}{2}\left(
-\hbar^2\frac{\partial^2}{\partial a^2}-\hbar^2\frac{\partial^2}{\partial b^2}
+a^2+b^2-2E
\right)\psi(a,b)=0.
\ee
Since $\hat{H}=\hat{H}_a+\hat{H}_b-E$, where $\hat{H}_a$ (resp. $\hat{H}_b$) is the nonrelativistic Hamiltonian for a simple harmonic oscillator in the variable $a$ (resp. $b$), this equation can be easily solved by using the basis that diagonalizes $\hat{H}_a$ and $\hat{H}_b$. Let
\be
\psi_n(a)\equiv\inner{a}{n}=\frac{1}{\sqrt{n!}}\,H_n(a)\,e^{-a^2/2\hbar}
\ee
be the normalized $n$th eigenfunction for the harmonic oscillator with eigenvalue $E_n=\hbar(n+1/2)$, where $H_n(a)$ is the $n$th
Hermite polynomial. Clearly, the function
\be
\psi_{n_a,n_b}(a,b):=\psi_{n_a}(a)\,\psi_{n_b}(b)\equiv\inner{a,b}{n_a,n_b}
\ee
solves \eqnref{eqn:qm constraint for double pendulum} if
\be
\hbar\left(n_a+n_b+1\right)=E,
\ee
which implies the quantum theory exists only if $E=\hbar(N+1)$ with $N\in\mathbb{Z}^+\cup\{0\}$.

Consequently, for a given $N$, the general solution of \eqnref{eqn:qm constraint for double pendulum} is given by
\be
\Psi(a,b)=\sum_{n=0}^{N} c_n\, \psi_n(a)\, \psi_{N-n}(b),
\ee
and thus the physical Hilbert space $\mathcal{H}$ is an $(N+1)$-dimensional proper subspace of $\mathcal{K}$ spanned by an orthonormal basis $\{\ket{n,N-n}\}_{n=0,\cdots,N}$.

The projector $\hat{P}:\mathcal{S}\rightarrow\mathcal{H}$ is a true projector as $\mathcal{H}$ is a proper subspace of $\mathcal{K}$ for the case that the spectrum of $\hat{H}$ is discrete. Obviously, $\hat{P}$ is given by
\be
\hat{P}=\sum_{n=0}^N \ket{n,N-n}\bra{n,N-n},
\ee
which can be obtained (up to an irrelevant overall factor) from \eqnref{eqn:projector}:
\ba
\int_0^{2\pi/\hbar}d\tau\, e^{-i\tau\hat{H}}
&\propto& \frac{1}{2\pi}\int_0^{2\pi}d\tau\sum_{n_a,n_b}\ket{n_a,n_b}
\,e^{-i\tau(n_a+n_b+1-E)}\bra{n_a,n_b}\nn\\
&=&\sum_{n_a,n_b}\delta_{n_a+n_b+1,E}\ket{n_a,n_b}\bra{n_a,n_b}
=\hat{P}.
\ea
Here, the integration range is so chosen because $\exp(-i\tau\hat{H})$ is periodic in $\tau$ with period $2\pi/\hbar$ if $E=\hbar(N+1)$.

The transition amplitudes are given by
\ba
W(a,b,a',b')&:=&\opelem{a,b}{\hat{P}}{a',b'}
=\sum_{n=0}^N \inner{a,b}{n,N-n}\inner{n,N-n}{a',b'}\nn\\
&=&\sum_{n=0}^N \frac{e^{-(a^2+b^2+a'^2+b'^2)/2\hbar}}{{n!(N-n)!}}\,
H_n(a)\,H_{N-n}(b)\,H_n(a')\,H_{N-n}(b')\,,
\ea
which is the probability density of measuring $(a,b)$, given $(a',b')$ measured at another instance. Furthermore, the probability of the quantum event $(n_a,n_b)$ given the quantum event $(n'_a,n'_b)$ is
\be
W[\psi_{n_a,n_b},\psi_{n'_a,n'_b}]:=\opelem{n_a,n_b}{\hat{P}}{n'_a,n'_b}
=\delta_{N,n_a+n_b}\delta_{n_a,n'_a}\delta_{n_b,n'_b}.
\ee

\subsection{More on the physical Hilbert space}\label{sec:physical Hilbert space}
The operator $\hat{P}\!: \mathcal{S} \rightarrow \mathcal{S}'$ maps an arbitrary element of $\mathcal{S}$ to its dual space $S'$.
If zero is in the continuous spectrum of $\hat{H}$, $\hat{P}$ maps $S$ to a larger space $S'$ and thus is \emph{not} really a projector. In this case, the physical state $\obra{\Psi_{\psi_\alpha}}$ mapped from $\ket{\psi_\alpha}$ is a tempered distribution. $\hat{P}$ becomes a true projector only if zero is a part of the discrete spectrum of $\hat{H}$ such as in the timeless double pendulum.

The construction in \eqnref{eqn:projector} is a special case for the \emph{group averaging} procedure \cite{Marolf:1995cn,Marolf:2000iq}, the idea of which is to averaging over all states along the gauge flow (generated by the constraint operator) to yield the physical solution which satisfies the constraint equation. In the special case, let $\ket{E}$ be the eigenstate of $\hat{H}$ with eigenvalue $E$, then schematically we have
\ba
\hat{H}\oket{\Psi_{\psi_\alpha}}&=&\int d\tau\, \hat{H} e^{-i\tau\hat{H}}\ket{\psi_\alpha}
=\int d\tau\! \int dE\, \hat{H} e^{-i\tau\hat{H}} \ket{E}\inner{E}{\psi_\alpha}\nn\\
&=&\int d\tau\! \int dE\, E\, e^{-i\tau E} \ket{E}\inner{E}{\psi_\alpha}
\propto\int dE\, \delta(E)\, E\ket{E}\inner{E}{\psi_\alpha}=0,
\ea
thus showing that $\hat{P}$ maps an arbitrary kinematical state $\ket{\psi_\alpha}$ to a physical state which satisfies the constraint equation \eqnref{eqn:qm constraint}. Furthermore, it can be easily shown that $\action{\Psi_{\psi_\alpha}}{\psi_\beta} =\action{\Psi_{\psi_\alpha}}{\psi_\beta'}$ if $\obra{\Psi_{\psi_\beta}}=\obra{\Psi_{\psi_\beta'}}$, and therefore the physical inner product in \eqnref{eqn:physical inner product} is well defined.

If there are multiple constraints, we have to solve the multiple constraint equations simultaneously:
\be
\hat{H}^i\ket{\psi}=0,
\qquad \text{for}\ i=1,\cdots,k.
\ee
In the simplest case that $[\hat{H}^i,\hat{H}^j]=0$ for all $i,j$, the projector can be easily constructed via
\be\label{eqn:projector multiple constraints}
\hat{P}=\int d\tau_1\cdots\int d\tau_k\, e^{-i\tau_i\hat{H}^i}
\ee
as a direct extension of \eqnref{eqn:projector}. In general, however, $\hat{H}^i$ do not commute, as classically the Poisson brackets $\{H^i,H^j\}$ vanish only \emph{weakly} (see \eqnref{eqn:first class} and \eqnref{eqn:first class 2}).

In the case that $\hat{H}^i$ do not commute but form a closed Lie algebra, i.e.,
\be
[\hat{H}^i,\hat{H}^j]={f^{ij}}_k\,\hat{H}^k
\ee
with ${f^{ij}}_k$ being constants, the exponentials of $\hat{H}^i$ form a Lie group $G$ and the physical state can be obtained by group averaging:
\be\label{eqn:group averaging}
\oket{\Psi_{\psi_\alpha}}=\int_G d\mu(\hat{U})\,\hat{U}\ket{\psi_\alpha},
\ee
where $d\mu$ is the Haar measure. It follows
\ba
\hat{U}'\oket{\Psi_{\psi_\alpha}}
&=&\int_G d\mu(\hat{U})\,\hat{U}'\hat{U}\ket{\psi_\alpha}
=\int_G d\mu(\hat{U}'^{-1}\hat{U}'')\,\hat{U}''\ket{\psi_\alpha}\nn\\
&=&\int_G d\mu(\hat{U}'')\,\hat{U}''\ket{\psi_\alpha}=\oket{\Psi_{\psi_\alpha}}
\ea
for any $\hat{U}'\in G$. The fact that $\oket{\Psi_{\psi_\alpha}}$ is invariant under any $\hat{U}'\in G$ implies that it is annihilated by the generators of $G$, namely, $\hat{H}^i\oket{\Psi_{\psi_\alpha}}=0$.
Furthermore, the physical inner product in \eqnref{eqn:physical inner product} is again well defined. (See \cite{Marolf:2000iq} for more details and subtleties.) The averaging in \eqnref{eqn:projector multiple constraints} is indeed a special case of \eqnref{eqn:group averaging}.

Generically, ${f^{ij}}_k$ are functions of $q^a$ and $p_a$ in \eqnref{eqn:first class 2}, and, correspondingly, $\hat{H}^i$ do not form a closed Lie algebra in the kinematical space $\mathcal{K}$. In this case, it is much more difficult to obtain the physical solutions and to construct the quantum theory which is free of quantum anomalies (see \cite{Thiemann:1996ay} for the issues of anomalies).

In \secref{sec:timeless path integral}, we will focus only on the case with a single Hamiltonian constraint, as our original motivation is to derive the timeless path integral in the absence of time. The trivial cases with multiple constraint are presented in \secref{sec:with multiple constraints}, while nontrivial cases are discussed in \secref{sec:discussion}.

\subsection{Remarks on measurements and collapse}\label{sec:measurements and collapse}
Imagine that a quantum system is measured by Alice and Bob at two different instances, yielding two outcomes corresponding to $\ket{\psi_\alpha}$ and $\ket{\psi_\beta}$, respectively. From the viewpoint of Alice, the physical state is collapsed to $\oket{\Psi_{\psi_\alpha}}$ by her measurement and Bob's measurement affirms her prediction. Bob, on the other hand, regards the physical state to be collapsed to $\oket{\Psi_{\psi_\beta}}$ by his measurement and predicts what Alice can measure. The striking puzzle arises: Who, Alice or Bob, causes the physical state to collapse in the first place?

In the timeless framework, it turns out to be an invalid question to ask who collapses the physical state \emph{first}, since we cannot make any sense of time. The seemingly puzzle is analogous to the Einstein-Podolsky-Rosen (EPR) paradox, in which a pair of entangled particles are measured separately by Alice and Bob. In the context of special relativity, if the two measurements are conducted at two spacetime events which are spacelike separated, the time-ordering of the two events can flip under a Lorentz boost and thus has no physical significance. Alice and Bob can both claim that the entangled state is collapsed by her/his measurement and thus have different knowledge about what the physical state should be, yet the predictions by Alice and Bob are consistent to each other. In our case, the measurement at an instance is analogous to the measurement on a single particle of the EPR pair; the kinematical state is analogous to the (local) state of a single particle; and the physical state is analogous to the (global) entangled state of the EPR pair. A complete knowledge (usually from measurement) about the local state will collapse the global state at once through the entanglement, which is analogous to the dynamics (or say, transition amplitudes) in our case. Consistency also holds in our case as $\action{\Psi_{\psi_\alpha}}{\psi_\beta}= \overline{\action{\Psi_{\psi_\beta}}{\psi_\alpha}}$\,. (See \cite{Laudisa} for more on the EPR paradox in the relational interpretation of quantum mechanics and also Section 5.6 of \cite{Rovelli:book} for more on the philosophical issues.)

If the system is deparametrizable, one can make sense of time-ordering with respect to the preferred time variable --- say, Alice performs her measurement before Bob. For Alice, the physical state is collapsed by her measurement and then she can predict Bob's measurement. What happens from the viewpoint of Bob? It turns out, for Bob, the physical state is collapsed by his measurement and he can \emph{retrodict} Alice's measurement. Alice's prediction and Bob's retrodiction again are consistent to each other. (See \cite{Marolf:2002ve} for more on consistency of prediction and retrodiction.) Analogously, in the EPR paradox, even if two measurements are conducted at two causally related events (and thus the time-ordering of the two events cannot be flipped), Alice's prediction and Bob's retrodiction lead to no inconsistency.\footnote{At first thought, it seems doubtful that (conventional) quantum mechanics can be used for retrodiction as well as for prediction, since this would imply that collapse by a measurement at present can affect states not only in the future but also in the past. It turns out this commits no violation against casuality and in fact is exactly what happens in the Wheeler's delayed-choice gedanken experiment \cite{Wheeler}, which has been confirmed by several experimental implementations \cite{Jacques}. The perplexity is closely related to the ``consistent histories'' interpretation of quantum mechanics \cite{Dowker}.}

As a side remark, exploiting further the close analogy between the EPR pair and the timeless formalism of relativistic quantum mechanics, one might be able to conceive an analog of the Bell's inequality, which would help to elaborate on the interpretational and conceptual issues of relativistic quantum mechanics at the level of thought experiments.

\section{Timeless path integral}\label{sec:timeless path integral}
The canonical formalism for relativistic quantum mechanics is described in \secref{sec:quantum theory}. All information of the quantum dynamics is encoded by the transition amplitudes \eqnref{eqn:W s s'}. In particular, by choosing $\ket{s}=\ket{q^a}$ and $\ket{s'}=\ket{q'^a}$, all physics can be obtained from the following transition amplitudes
\ba
W(q^a,q'^a)=\opelem{q^a}{\hat{P}}{q'^a}
\sim\int d\tau \, \opelem{q^a}{e^{-i\tau \hat{H}}}{q'^a}.
\ea
From now on, we will use the notation $\sim$ to denote the equality up to an overall constant factor which has no physical significance, as any overall constant is canceled out in the numerator and denominator in \eqnref{eqn:P s s'}.

As a special case of group averaging, the integration range of $\tau$ is taken to be a compact interval if $\exp(-i\tau\hat{H})$ forms a compact Lie group $U(1)$ (timeless double pendulum is an example) and it is taken to be $(-\infty,\infty)$ if the group of $\exp(-i\tau\hat{H})$ is noncompact. For the case that $\exp(-i\tau\hat{H})$ gives $U(1)$, we can unwrap $U(1)$ to its covering space $\mathbb{R}$ and correspondingly integrate $\tau$ over $(-\infty,\infty)$. The unwrapping only gives rise to an overall multiplicative factor (which is divergent if not properly regulated). Therefore, in any case, up to an irrelevant overall factor, transition amplitudes can be computed by
\be\label{eqn:W q q'}
W(q^a,q'^a)
\sim \int_{-\infty}^\infty d\tau \, \opelem{q^a}{e^{-i\tau \hat{H}}}{q'^a},
\ee
where $\opelem{q^a}{e^{-i\tau \hat{H}}}{q'^a}$ can be thought as the transition amplitude for a kinematical state $\ket{q'^a}$ to ``evolve' to the state $\ket{q^a}$ by the ``parameter time'' $\tau$. Equation \eqnref{eqn:W q q'} sums over $\opelem{q^a}{e^{-i\tau \hat{H}}}{q'^a}$ for all possible values of $\tau$, suggesting that $W(q^a,q'^a)$ is intrinsically timeless as the parameter time $\tau$ has no essential physical significance.

Rigorously, the integration should be regularized via
\be\label{eqn:regularization}
W(q^a,q'^a)\sim
\lim_{M\rightarrow\infty}
\frac{\int_{-M}^M d\tau \opelem{q^a}{e^{-i\tau \hat{H}}}{q'^a}}
{\int_{-M}^M d\tau},
\ee
as a cut-off $M$ is introduced to regulate the integral and the irrelevant overall factor to be finite. As we will see, the variable $\tau$ corresponds to the parametrization of curves in the path integral and integrating over all $\tau$ indicates that the parametrization of curves has no physical significance. The above regularization scheme is physically well justified, as it cuts off the curves in the path integral which are ``too wild'' (noncompact curves), given the endpoints $q'^a$ and $q^a$ fixed.

In the following, starting from \eqnref{eqn:W q q'}, we will first derive the timeless path integral for the case of a single Hamiltonian constraint and then investigate it in more detail. In the end, we will study the path integral with multiple relativistic constraints which commute mutually.

\subsection{General structure}\label{sec:path integral general structure}
For a given $\tau$, let us introduce a \emph{parametrization sequence}: $\tau_0=0,\, \tau_1,\,\tau_2,\cdots\!,\tau_{N-1},\,\tau_N=\tau$ with $\tau_n\in\mathbb{R}$, and define $\Delta\tau_n:=\tau_n-\tau_{n-1}$. The conditions on the endpoints ($\tau_0=0$ and $\tau_N=\tau$) correspond to $\sum_{n=1}^N\Delta\tau_n=\tau$. The \emph{mesh} of the parameter sequence is defined to be $\max_{n=1,\cdots,N}\{\left|\Delta\tau_n\right|\}$. The parameter sequence is said to be fine enough if its mesh is smaller than a given small number $\epsilon$.\footnote{\label{foot:parametrization}Let $X$ be a topological space and $s\in[0,1]$. A continuous map $\gamma:\,s\mapsto\gamma(s)\in X$ is called a \emph{path} with an initial point $s(0)=x_0$ and an end point $s(1)=x_1$. The image of $\gamma$ is called a \emph{curve}, which can be reparameterized with respect to a new variable $\tau$ as $\gamma:\,\tau\mapsto\gamma(\tau)$ by introducing an arbitrary continuous function $\tau:\,s\mapsto\tau(s)\in\mathbb{R}$. The parametrization sequence $\tau_0=0,\, \tau_1,\,\tau_2,\cdots\!,\tau_{N-1},\,\tau_N=\tau$ can be viewed as a discrete approximation for the reparametrization function $\tau(s)$ with $\tau(s=0)=0$ and $\tau(s=1)=\tau$ if we identify $\tau_n=\tau(n/N)$. For the case that $\tau(s)$ is injective, the parametrization sequence is ordered (i.e.\ $0=\tau_0<\tau_1<\cdots<\tau_{N-1}<\tau_N=\tau$ and $\Delta\tau_n>0$ if $\tau>0$) and called a \emph{partition} of the interval $[0,\tau]$, which is used to define the Riemann integral as the continuous limit: $\int_0^\tau f(\tau')d\tau'=\lim_{\mathrm{mesh}\rightarrow0}\sum_{n=0}^{N-1}f(\tau_n)\Delta\tau_{n+1}$. In the timeless formulation of relativistic mechanics, a dynamical solution is an \emph{unparameterized} curve in $\Omega$ and its parametrization has no physical significance. In order to exploit the timeless feature, we should keep the parametrizations generic and not restrict ourselves to injective ones. Correspondingly, the partition is generalized to a parametrization sequence and the Riemann integral is generalized to the Riemann-Stieltjes integral as $\int_0^1 f(s)d\tau(s)=\lim_{\mathrm{mesh}\rightarrow0}\sum_{n=0}^{N-1}f(n/N)\Delta\tau_{n+1}$, which is well defined even if $\tau(s)$ is not injective.}

As $\tau$ is fixed now, identifying $q^a=q_N^a$ and $q'^a=q_0^a$, and using $\sum_{n=1}^N\Delta\tau_n=\tau$, we can rewrite $\opelem{q^a}{e^{-i\tau \hat{H}}}{q'^a}$ as
\ba\label{eqn:kernel}
&&\opelem{q^a}{e^{-i\tau \hat{H}}}{q'^a}
\equiv \opelem{q_N^a}{e^{-i\Delta\tau_N\hat{H}}\, e^{-i\Delta\tau_{N-1}\hat{H}} \cdots e^{-i\Delta\tau_1\hat{H}}}{q_0^a}\nn\\
&=&
\left(\prod_{n=1}^{N-1}\int d^dq_n^a\right)
\opelem{q_N^a}{e^{-i\Delta\tau_N \hat{H}}}{q_{N-1}^a}
\opelem{q_{N-1}^a}{e^{-i\Delta\tau_{N-1} \hat{H}}}{q_{N-2}^a}
\cdots
\opelem{q_1^a}{e^{-i\Delta\tau_1 \hat{H}}}{q_0^a},
\ea
where we have inserted $N-1$ times the completeness relation
\be
\int d^dq^a \, \ket{q^a}\bra{q^a}
:= \int dq^1\cdots dq^d \, \ket{q^1,\cdots,q^d}\bra{q^1,\cdots,q^d}\,.
\ee
For a given arbitrary small number $\epsilon$, by increasing $N$, we can always make the parameter sequence fine enough such that $\mathrm{mesh}\{\tau_i\}\leq|\tau|/N\leq\epsilon$.\footnote{More rigorously, for a given $\epsilon$, the large number $N$ should be chosen to satisfy $\mathrm{mesh}\{\tau_i\}\leq|\tau|/N\leq M/N\leq\epsilon$, where $M$ is the cut-off regulator defined in \eqnref{eqn:regularization}, so that the $\mathcal{O}(\epsilon^2)$ term in \eqnref{eqn:approximation} can be dropped for any value of $|\tau|$. In the end, we have to integrate \eqnref{eqn:discrete path integral for kernel} over all possible values of $\tau$ to obtain $W(q^a,q'^a)$, and the regularization is essential to keep the $\mathcal{O}(\epsilon^2)$ terms under control for arbitrary values of $\tau$.}
Consequently, we can approximate each $\opelem{q_{n+1}^a}{e^{-i\Delta\tau_{n+1} \hat{H}}}{q_n^a}$ to the first order in $\epsilon$ as
\be\label{eqn:step evolution}
\opelem{q_{n+1}^a}{e^{-i\Delta\tau_{n+1} \hat{H}}}{q_{n}^a}
=\opelem{q_{n+1}^a}{1-i\Delta\tau_{n+1}\hat{H}(\hat{q}^a,\hat{p}_a)}{q_{n}^a}
+\mathcal{O}(\epsilon^2).
\ee

For the generic case that the Hamiltonian operator $\hat{H}$ is a polynomial of $\hat{q}^a$ and $\hat{p}_a$ and is \emph{Weyl ordered}, with the use of the completeness relation for the momenta
\be
\int \frac{d^dp_a}{(2\pi\hbar)^d} \, \ket{p_a}\bra{p_a}
:= \int \frac{dp_1\cdots dp_d}{(2\pi\hbar)^d}
 \, \ket{p_1,\cdots,p_d}\bra{p_1,\cdots,p_d},
\ee
it can be shown that
\be\label{eqn:q H q}
\opelem{q^a}{\hat{H}(\hat{q}^q,\hat{p}_a)}{q'^a}
=\int\frac{d^dp_a}{(2\pi\hbar)^d}\,
\exp\left[\frac{i}{\hbar}\,p_a(q^a-q'^a)\right]
H\left(\frac{q^a+q'^a}{2},p_a\right).
\ee
(See Exercise~11.2 in \cite{Greiner:book} for the proof.)
Applying \eqnref{eqn:q H q} to \eqnref{eqn:step evolution}, we have
\ba\label{eqn:approximation}
\opelem{q_{n+1}^a}{e^{-i\Delta\tau_{n+1} \hat{H}}}{q_{n}^a}
&=&\int\frac{d^dp_{na}}{(2\pi\hbar)^d}\
e^{ip_{na}(q_{n+1}^a-q_n^a)/\hbar}
\left[1-i\Delta\tau_{n+1}H\left(\frac{q_{n+1}^a\!+\!q_n^a}{2},p_{na}\right)\right] +\mathcal{O}(\epsilon^2)\nn\\
&=&\int\frac{d^dp_{na}}{(2\pi\hbar)^d}\
e^{ip_{na}\Delta q_n^a/\hbar}\
e^{-i\Delta\tau_{n+1}H(\bar{q}_n^a,\,p_{na})}+\mathcal{O}(\epsilon^2),
\ea
where we define $\bar{q}_n^a:=(q_{n+1}^a+q_n^a)/2$ and $\Delta q_n^a:=q_{n+1}^a-q_n^a$.

Making the parametrization sequence finer and finer (by decreasing $\epsilon$ or equivalently by increasing $N$) and at the end going to the limit $\epsilon\rightarrow0$ or $N\rightarrow\infty$, we can cast \eqnref{eqn:kernel} as
\ba\label{eqn:discrete path integral for kernel}
\opelem{q^a}{e^{-i\tau \hat{H}}}{q'^a}
&=&\lim_{N\rightarrow\infty}
\left(\prod_{n=1}^{N-1}\int d^dq_n^a\right)
\left(\prod_{n=0}^{N-1}\int\! \frac{d^dp_{na}}{(2\pi\hbar)^d}\right)
\exp\left(\frac{i}{\hbar}\sum_{n=0}^{N-1}p_{na}\Delta q_n^a\right)\nn\\
&&\quad\times
\exp\left(-i\sum_{n=0}^{N-1}\Delta\tau_{n+1}H(\bar{q}_n^a,p_{na})\right).
\ea
In the limit $N\rightarrow\infty$, the points $q_n$ and $p_n$ can be viewed as the sampled points of a continuous curve in $\Omega=T^*\mathcal{C}$ given by $\tilde{\gamma}(\tau')=(q^a(\tau'),p_a(\tau'))$, which is parameterized by $\tau'$ and with the endpoints projected to $\mathcal{C}$ fixed by $q^a(\tau'=0)=q'^a$ and $q^a(\tau'=\tau)=q^a$. That is, $q_n$ and $p_n$ are the sampled points of $\tilde{\gamma}$ as $q_n^a=q^a(\tau_n)$ and $p_{na}=p_a(\tau_n)$. In the treatment of functional integral, it is customary to introduce the special notations for path integrals:
\begin{subequations}
\ba
\prod_{n=1}^{N-1}\int d^dq_n^a \quad &\rightarrow& \quad \int\mathcal{D}q^a,\\
\prod_{n=0}^{N-1}\int\! \frac{d^dp_{na}}{(2\pi\hbar)^d} \quad &\rightarrow& \quad \int\mathcal{D}p_a.
\ea
\end{subequations}
Meanwhile, in the continuous limit ($N\rightarrow\infty$), the finite sums appearing in the exponents in \eqnref{eqn:discrete path integral for kernel} also converge to the integrals:
\be\label{eqn:int p dq}
\frac{i}{\hbar}\sum_{n=0}^{N-1}p_{na}\Delta q_n^a
\quad\rightarrow\quad
\frac{i}{\hbar}\int_{\tilde{\gamma}} p_adq^a
\equiv\frac{i}{\hbar}\int_{\tilde{\gamma}} \left(p_a\frac{dq^a}{d\tau'}\right) d\tau'
\ee
and
\be\label{eqn:int H dtau}
-i\sum_{n=0}^{N-1}\Delta\tau_{n+1}H(p_{na},\,\bar{q}_n^a)
\quad\rightarrow\quad
-i\int_{\tilde{\gamma}} H(q^a(\tau'),p_a(\tau'))\,d\tau'.
\ee
Note that the continuous limit above is defined via the Riemann-Stieltjes integral as an extension of the Riemann integral (see \footref{foot:parametrization}).
With the new notations, \eqnref{eqn:discrete path integral for kernel} can be written in a concise form:
\be\label{eqn:continuous path integral for kernel}
\opelem{q^a}{e^{-i\tau \hat{H}}}{q'^a}=
\int\mathcal{D}q^a \int\mathcal{D}p_a\
\exp\left(\frac{i}{\hbar}\int_{\tilde{\gamma}}p_adq^a\right)
\exp\left(-i\int_{\tilde{\gamma}} H(q^a(\tau'),p_a(\tau'))\,d\tau'\right).
\ee

It is remarkable to note that (up to the factor $i/\hbar$) the continuous limit in \eqnref{eqn:int p dq} is simply the line integral of the one-form $\tilde{\theta}=p_adq^a$ over the curve $\tilde{\gamma}$, identical to \eqnref{eqn:cl action}, and is independent of the parametrization of $\tau$. On the other hand, the integral in \eqnref{eqn:int H dtau} depends on the parametrization of $\tau$. Thus, to compute $W(q^a,q'^a))$ in \eqnref{eqn:W q q'}, the integration over $\tau$ only hits the second exponential in \eqnref{eqn:continuous path integral for kernel} and the first exponential simply factors out. The integration of the second exponential over $\tau$ yields
\ba\label{eqn:delta function}
&&\int_{-\infty}^\infty d\tau \exp\left(-i\int_{\tilde{\gamma}} H(q^a(\tau'),p_a(\tau'))\,d\tau'\right)\nn\\
&=&\int_{-\infty}^\infty d\tau \exp\left(-i\tau\int_{\tilde{\gamma}} H(q^a(\bar{\tau}),p_a(\bar{\tau}))\,d\bar{\tau}\right)
=\delta\left(\int_{\tilde{\gamma}} H(q^a(\bar{\tau}),p_a(\bar{\tau}))\,d\bar{\tau}\right),
\ea
where we have rescaled the parametrization $\tau'$ to $\bar{\tau}=\tau'/\tau$ so that the endpoints now read as $q'^a=q^a(\bar{\tau}=0)$ and $q^a=q^a(\bar{\tau}=1)$.\footnote{\label{foot:removing regulator}In the expression of \eqnref{eqn:delta function}, we have removed the cut-off regulator (i.e.\ the limit $M\rightarrow\infty$ has been taken). More rigorously, we have removed the regulator \emph{before} the limit $N\rightarrow\infty$ is taken. The Dirac delta function in \eqnref{eqn:delta function} would have been a \emph{nascent delta function} if the regulator had not been removed.} The appearance of the Dirac delta function indicates that only the paths which satisfy $\int_{\tilde{\gamma}}H(\bar{\tau})d\bar{\tau}=0$ will contribute to the path integral for $W(q^a,q'^a)$. The condition $\int_{\tilde{\gamma}}H(\bar{\tau})d\bar{\tau}=0$ is, however, still not geometrical, since $\bar{\tau}$ can be further reparameterized to $\bar{\tau}'=\bar{\tau}'(\bar{\tau})$ to yield $\int_{\tilde{\gamma}}H(\bar{\tau}')d\bar{\tau}'\neq0$ even with the initial and final values fixed, i.e., $\bar{\tau}'(\bar{\tau}=0)=0$ and $\bar{\tau}'(\bar{\tau}=1)=1$. On the other hand, $W(q^a,q'^a)$ cast in \eqnref{eqn:W q q'} has no dependence on the parametrization whatsoever, which implies that, in the continuous limit, the contribution of a path $\tilde{\gamma}$ satisfying the condition $\int_{\tilde{\gamma}}H(\bar{\tau})d\bar{\tau}=0$ for a specific (rescaled) parametrization $\bar{\tau}$ is somehow exactly canceled by that of another path satisfying the same condition. In the end, only the paths restricted to the constraint surface (i.e., $\tilde{\gamma}\in\Sigma$, or equivalently $H(\tau')=0$ for all $\tau'$ along the path) contribute to the path integral for $W(q^a,q'^a)$. The constraint $\tilde{\gamma}\in\Sigma$ is now geometrical.

How the aforementioned cancelation takes place is obscure. To elucidate this point, we exploit the fact that $W(q^a,q'^a)$ is independent of the parametrization and play the trick by averaging over all possible parametrizations. That is, up to an overall factor of no physical significance, we can recast $W(q^a,q'^a)$ by summing over different parametrizations as follows:
\ba\label{eqn:path integral 1}
&&W(q^a,q'^a)
\sim \int d\tau \int \left[\mathcal{D}\Delta\tau\right]_{\sum\!\Delta\tau_n=\tau}
\opelem{q^a}{e^{i\tau \hat{H}}}{q'^a}\\
&\sim& \int d\tau \int\left[\mathcal{D}\Delta\tau\right]_{\sum\!\Delta\tau_n=\tau} \int\mathcal{D}q^a \int\mathcal{D}p_a\
\exp\left(\frac{i}{\hbar}\int_{\tilde{\gamma}}p_adq^a\right)
\exp\left(-i\sum_{n=0}^{N-1}\Delta\tau_{n+1}H(\bar{q}_n^a,p_{na})\right),\nn
\ea
where the notation $\int\left[\mathcal{D}\Delta\tau\right]_{\sum\!\Delta\tau_n=\tau}$ is a shorthand for
\be
\underbrace{\int_{-\tau/N}^{\tau/N} d\Delta\tau_1 \int_{-\tau/N}^{\tau/N} d\Delta\tau_2 \cdots \int_{-\tau/N}^{\tau/N} d\Delta\tau_N}_{\sum_{n=1}^{N}\Delta\tau_n=\tau}
\quad\rightarrow\quad
\int\left[\mathcal{D}\Delta\tau\right]_{\sum\!\Delta\tau_n=\tau},
\ee
which sums over all fine enough  (namely, $\mathrm{mesh}\{\tau_i\}\leq|\tau|/N$) parametrization sequences for a given $\tau$. It is easy to show that
\be
\int_{-\infty}^\infty d\tau
\underbrace{\int_{-\tau/N}^{\tau/N} d\Delta\tau_1 \int_{-\tau/N}^{\tau/N} d\Delta\tau_2 \cdots \int_{-\tau/N}^{\tau/N} d\Delta\tau_N}_{\sum_{n=1}^{N}\Delta\tau_n=\tau}
=\prod_{n=0}^{N-1}\int_{-\infty}^{\infty} d\Delta\tau_{n+1},
\ee
when the cut-off regulator $M$ is removed (also see \footref{foot:removing regulator}).
Consequently, for a given arbitrary parametrization $\tau'$, renaming the varying $\Delta\tau_n$ as $\Delta\tau_n=\hbar^{-1} N_n\Delta\tau'_n$, we can rewrite \eqnref{eqn:path integral 1} as
\ba\label{eqn:path integral 2}
&&W(q^a,q'^a)\\
&\sim& \int\mathcal{D}q^a \int\mathcal{D}p_a \int\mathcal{D}N\
\exp\left(\frac{i}{\hbar}\int_{\tilde{\gamma}}p_adq^a\right)
\exp\left(-\frac{i}{\hbar}\sum_{n=0}^{N-1}
\Delta\tau'_{n+1}N_{n+1}H(\bar{q}_n^a,p_{na})\right),\nn
\ea
where we introduce the notation
\be
\prod_{n=0}^{N-1}\int_{-\infty}^\infty dN_{n+1} \quad \rightarrow \quad \int\mathcal{D}N.
\ee
Again, in the continuous limit, the finite sum converges to the Riemann-Stieltjes integral:
\be
-\frac{i}{\hbar}\sum_{n=0}^{N-1}\Delta\tau'_{n+1}N_{n+1}H(\bar{q}_n^a,p_{na})
\quad\rightarrow\quad
-\frac{i}{\hbar}\int_{\tilde{\gamma}}
N(\tau')H(q^a(\tau'),p_a(\tau'))d\tau',
\ee
and \eqnref{eqn:path integral 2} can be neatly written as the \emph{path integral}:
\ba\label{eqn:path integral 3}
W(q^a,q'^a)
&\sim& \int\mathcal{D}q^a \int\mathcal{D}p_a \int\mathcal{D}N\
\exp\left[\frac{i}{\hbar}
\left(\int_{\tilde{\gamma}}p_adq^a
-\int_{\tilde{\gamma}}N(\tau')H d\tau'\right)
\right]\nn\\
&\equiv& \int\mathcal{D}q^a \int\mathcal{D}p_a \int\mathcal{D}N\
\exp\left[\frac{i}{\hbar}
\int_{\tilde{\gamma}}\left(p_a\frac{dq^a}{d\tau'}
-N(\tau')H\right)d\tau'\right].
\ea
Integration over $N$ can be carried out to obtain the delta functional:
\be
\int\mathcal{D}N \exp\left(\frac{i}{\hbar}\int N(\tau')H d\tau'\right)\sim\delta[H]
\equiv \lim_{N\rightarrow\infty}\prod_{n=0}^{N-1} \delta(H(\bar{q}^a_n,p_{an})),
\ee
and thus the path integral \eqnref{eqn:path integral 3} can be written in an alternative form as
\be\label{eqn:path integral 4}
W(q^a,q'^a)\sim
\int\mathcal{D}q^a \int\mathcal{D}p_a\ \delta[H]
\exp\left[\frac{i}{\hbar}
\int_{\tilde{\gamma}}p_adq^a\right],
\ee
where insertion of the delta functional $\delta[H]$ confines the path to be in the constraint surface (i.e.\ $\tilde{\gamma}\in\Sigma$). Note that the phase in the exponent in \eqnref{eqn:path integral 4} is identical to the classical action defined in \eqnref{eqn:cl action} (divided by $\hbar$) and that in \eqnref{eqn:path integral 3} is identical to the classical action in \eqnref{eqn:cl action with N} with $k=1$. Therefore, each path in $\Sigma$ contributes with a phase, which is the classical action divided by $\hbar$.

The path integral formalism is intuitively appealing. It gives us an intuitive picture about the transition amplitudes: $W(q^a,q'^a)$ is described as the sum, with the weight $\exp(iS/\hbar)$ (where $S$ is the classical action of $\tilde{\gamma}$), over all arbitrary paths $\tilde{\gamma}$ which are restricted to $\Sigma$ and whose projection $\gamma$ to $\mathcal{C}$ connect $q'^a$ and $q^a$. None of $q^a$ is restricted to be monotonic along the paths, and in this sense the formulation is called \emph{timeless} path integral. The parametrization for the paths has no physical significance as can be seen in the expression of \eqnref{eqn:path integral 4}, which is completely geometrical and independent of parametrizations. On the other hand, the continuum notation of \eqnref{eqn:path integral 3} is really a schematic for the discretized version:
\begin{subequations}
\ba
\label{eqn:path integral 5 a}
W(q^a,q'^a) &\sim& \lim_{N\rightarrow\infty}
\prod_{n=1}^{N-1}\int d^dq_n^a
\prod_{n=0}^{N-1}\int\! \frac{d^dp_{na}}{(2\pi\hbar)^d}
\prod_{n=0}^{N-1}\int dN_{n+1}\nn\\
&&\qquad\qquad\times
\exp\left(-\frac{i}{\hbar}\sum_{n=0}^{N-1}
\Delta\tau'_{n+1}N_{n+1}H(\bar{q}_n^a,p_{na})\right)\\
\label{eqn:path integral 5 b}
&\sim&
\lim_{N\rightarrow\infty}
\prod_{n=1}^{N-1}\int d^dq_n^a
\prod_{n=0}^{N-1}\int\! \frac{d^dp_{na}}{(2\pi\hbar)^d}
\prod_{n=0}^{N-1}\int dN_{n+1}\nn\\
&&\qquad\qquad\times
\exp\left(-\frac{i}{\hbar}\sum_{n=0}^{N-1}
N_{n+1}H(\bar{q}_n^a,p_{na})\right),
\ea
\end{subequations}
where $\Delta\tau'_n$ in \eqnref{eqn:path integral 5 a} is absorbed to $N_n$ in \eqnref{eqn:path integral 5 b} and this only results in an irrelevant overall factor. The expression of \eqnref{eqn:path integral 5 b} is explicitly independent of parametrizations.\footnote{Perhaps, more appropriately, the ``timeless path integral'' should be renamed ``timeless `curve' integral'', as in the rigorous terminology, a \emph{curve} is defined as the unparameterized image of a \emph{path}, which is specified by a parameter. However, we keep the name of ``path integral'' to conform to the conventional nomenclature.}

The contributing paths in the path integral can be very ``wild'' --- not necessarily smooth or even continuous. This calls into question whether the path integral can achieve convergence. We do not attempt to present a rigorous derivation here but refer to \cite{Simon:book} for the legitimacy issues and subtleties of the path integral.

Each path in $\Sigma$ contributes with a different phase, and the contributions from the paths essentially cancel one another through destructive interference until we come near the stationary solution. As a result, most contributions come from the paths close to the stationary solution. The stationary solution can be obtained by taking the functional variations on \eqnref{eqn:path integral 3} with respect to $N$, $p_a$ and $q^a$, which yield the classical Hamiltonian constraint \eqnref{eqn:cl constraint} and the Hamilton equations \eqnref{eqn:Hamilton eqs}; that is, the stationary solution coincides with the classical solution. Provided that the action for the classical solution is much greater than $\hbar$, i.e.\ $S[\tilde{\gamma}_i]\gg\hbar$, the stationary phase approximation (see \appref{app:stationary approximation}) yields
\be\label{eqn:W approx}
W(q^a,q'^a)\approx \sum_i \,\zeta_{\tilde{\gamma}_i}\, e^{\frac{i}{\hbar}S[\tilde{\gamma_i}]},
\ee
where $\tilde{\gamma_i}$ are the classical solutions which connect $q'^a$ and $q^a$,\footnote{Generally, there could be multiple classical solutions connecting $q'^a$ and $q^a$ (as in the case of the timeless double pendulum), especially when the system is not deparametrizable.} and $\zeta_{\tilde{\gamma}_i}$ are the weight factors for each classical solution $\tilde{\gamma}_i$, which are proportional to the decoherence width of phases of the nearby trajectories around $\tilde{\gamma}_i$.\footnote{There could be possible experimental tests of the weight factors in mesoscopic phenomena or optical/electron diffractions/interferences.}
By expanding the paths around the classical solutions, various semiclassical approximation methods in (conventional) path integral approaches can be easily carried over to the timeless path integral.

\subsection{Deparametrizable systems as a special case}\label{sec:deparametrizable systems}
If the Hamiltonian happens to be deparametrizable, the classical Hamiltonian is in the form of \eqnref{eqn:nonrel H}, and the path integral \eqnref{eqn:path integral 4} reads as
\begin{subequations}\label{eqn:path integral deparametrizable}
\ba
W(q^a,q'^a)&\sim&
\int\mathcal{D}t\int\mathcal{D}q^i \int\mathcal{D}p_t\int\mathcal{D}p_i\ \delta[p_t+H_0]
\exp\left[\frac{i}{\hbar}
\int_{\tilde{\gamma}}\left(p_tdt+p_idq^i\right)\right]\nn\\
&=&\int\mathcal{D}t\int\mathcal{D}q^i \int\mathcal{D}p_i\,
\exp\left[\frac{i}{\hbar}
\int_{\tilde{\gamma}}\left(p_i\frac{dq^i}{dt}-H_0\right)dt\right]\\
&\equiv&
\lim_{N\rightarrow\infty}
\left(\prod_{n=1}^{N-1}\int dt_n\right)
\left(\prod_{n=1}^{N-1}\int d^{d-1}q_n^i\right)
\left(\prod_{n=0}^{N-1}\int\! \frac{d^{d-1}p_{ni}}{(2\pi\hbar)^{d-1}}\right)\nn\\
&&\quad\times\exp\left[\frac{i}{\hbar}\sum_{n=0}^{N-1}
\left(p_{ni}\frac{\Delta q_n^i}{\Delta t_n}-H_0(\bar{q}^i_n,p_{in};t_n)\right)\Delta t_n
\right].
\ea
\end{subequations}

If the system is strictly deparametrizable, i.e., $[\hat{H}_0(t_1),\hat{H}_0(t_2)]=0$, the transition amplitude for conventional quantum mechanics, denoted as $G(q^i,t;q'^i,t')$, is given by \eqnref{eqn:conventional path integral}. Note that, for given arbitrary $t_1,t_2,\cdots,t_N$, the integrand for $\int\mathcal{D}t$ in \eqnref{eqn:path integral deparametrizable} is formally identical to $G(q^i,t;q'^i,t')$ given in \eqnref{eqn:conventional path integral} (with $t_n$ replaced by $\tau_n$), and thus we have
\be\label{eqn:W and G}
W(q^a,q'^a)\sim\int\mathcal{D}t\, G(q^i,t;q'^i,t') \sim G(q^i,t;q'^i,t'),
\ee
where $\int\mathcal{D}t$ simply factors out as an irrelevant overall factor. This implies that $W(t,q^i,t',q'^i)$ and $G(t,q^i;t',q'^i)$ are identical to each other (up to an irrelevant normalization factor for $W$) for strictly deparametrizable systems as commented in \secref{sec:remarks on deparametrizable systems}.

For non-strictly deparametrizable systems, i.e., $[\hat{H}_0(t_1),\hat{H}_0(t_2)]\neq0$, on the other hand, we do not have \eqnref{eqn:W and G}, because while \eqnref{eqn:path integral deparametrizable} sums over all possible paths $\tilde{\gamma}(\tau)=(q^a(\tau),p_a(\tau))=(t(\tau),q^i(\tau),p_t(\tau),p_i(\tau))$ which can move forward and backward in $t$, \eqnref{eqn:conventional path integral} now sums over only the paths $\tilde{\gamma}_0(t)=(q^i(t),p_i(t))$ which are monotonic in $t$ as the time-ordered condition \eqnref{eqn:ordered tau} has to be imposed for the systems in which $[\hat{H}_0(\hat{q}^i,\hat{p}_i;t_1),\hat{H}_0(\hat{q}^i,\hat{p}_i;t_2)]\neq0$.
The difference is profound and shows that relativistic quantum mechanics and conventional quantum mechanics are different both at the level of kinematics and at the level of dynamics if the system is non-strictly deparametrizable, as already commented in the end of \secref{sec:remarks on deparametrizable systems}.

However, for most situations, provided that the action for the classical solution is much greater than $\hbar$, we have the good approximation \eqnref{eqn:W approx} and only the paths in the vicinity of the classical solution are important. Meanwhile, as discussed in \secref{sec:nonrelativistic mechanics}, the classical solution for a deparametrizable system is always monotonic in $t$. Thus, for non-strictly deparametrizable systems, it is a good approximation in \eqnref{eqn:path integral deparametrizable} to sum over only the paths which are not too deviated from the classical solution and are monotonic in $t$. In this approximation, \eqnref{eqn:path integral deparametrizable} reduces to the conventional path integral \eqnref{eqn:conventional path integral} as $\int\mathcal{D}t$ factors out as an irrelevant overall factor. Therefore, the conventional path integral, although not equivalent to, is a good approximation for the timeless path integral for non-strictly deparametrizable systems. Further research is needed to investigate when the approximation remains good and when it fails. This issue is closely related to the composition laws of relativistic quantum mechanics studied in \cite{Halliwell:1992nj} and the main idea of \cite{Halliwell:2012nj} that the complication with the quantum Zeno effect in the conventional path integral should be avoided by ``softening'' the restriction on paths in a manner which gives rise to coarse-graining in time scale (also see \cite{Halliwell:2009rw} for the issues of probability distributions in the context of the decoherent histories approach to quantum theory).

\subsection{Timeless Feynman's path integral}
Consider the special case that the classical Hamiltonian is given in the form of \eqnref{eqn:H special form} and the Hamiltonian operator is Weyl ordered. As the Hamiltonian is a quadratic polynomial in $p_a$, the path integral over $\mathcal{D}p_a$ in \eqnref{eqn:path integral 3} can be integrated out. That is, in the expression:\footnote{In this subsection, the repeated index $a$ is not summed unless $\sum_a$ is explicitly used.}
\ba
&&W(q^a,q'^a)\\
&\sim& \int\mathcal{D}q^a \int\mathcal{D}N
\prod_{n=0}^{N-1}\int\! \frac{d^dp_n}{(2\pi\hbar)^d}\
\exp\left[\frac{i}{\hbar}\sum_{n=0}^{N-1}
\left(\sum_a p_{na}\Delta q_n^a-\Delta\tau'_{n+1}N_{n+1}H(\bar{q}_n^a,p_{na})\right)
\right],\nn
\ea
the integration over each $p_{na}$ can be explicitly carried out:
\ba
&&\int_{-\infty}^\infty dp_{na}
\exp\left(\frac{i}{\hbar}
\left[
p_{na}\Delta q_n^a-\Delta\tau'_{n+1}N_{n+1}
\left(\alpha_ap_{na}^2 + \beta_ap_{na}\bar{q}_n^a+
\gamma_ap_{na}\right)
\right]\right)\nn\\
&\propto&
\frac{1}{\sqrt{\rule{0mm}{3.5mm}N_{n+1}}}\,
\exp \left(\frac{i}{\hbar}\frac{\Delta\tau'_{n+1}N_{n+1}}{4\alpha_a}
\left[\frac{\Delta q_n^a}{\Delta\tau'_{n+1}N_{n+1}}
-\beta_a\bar{q}_n^a-\gamma_a\right]^2 \right)
\ea
by the Gaussian integral $\int_{-\infty}^\infty dx\, e^{-\alpha x^2+\beta x}=(\pi/\alpha)^{1/2}e^{\beta^2/4\alpha}$.
Noting that $dN_{n+1}/\sqrt{\rule{0mm}{3.5mm}N_{n+1}}=2\,d\sqrt{\rule{0mm}{3.5mm}N_{n+1}}$\, and introducing the shorthand notation:
\be
\prod_{n=0}^{N-1}\int_{-\infty}^\infty d\sqrt{N_{n+1}} \quad \rightarrow \quad \int\mathcal{D}\sqrt{N},
\ee
we then have
\ba
W(q^a,q'^a)&\sim& \int\mathcal{D}q^a \int\mathcal{D}\sqrt{N}\\
&&\times
\exp\left[\frac{i}{\hbar}\sum_{n=0}^{N-1}
\left(\sum_a \frac{N_{n+1}}{4\alpha_a}
\left[\frac{\Delta q_n^a}{\Delta\tau'_{n+1}N_{n+1}}
-\beta_a\bar{q}_n^a-\gamma_a\right]^2-N_{n+1}V(\bar{q}_n^a)\right)\Delta\tau'_{n+1}
\right],\nn
\ea
which written in the continuous form reads as
\be\label{eqn:Feynman's path integral}
W(q^a,q'^a)
\sim \int\mathcal{D}q^a \int\mathcal{D}\sqrt{N}
\exp\frac{i}{\hbar}\int_\gamma d\tau'
\left(\sum_a\frac{N}{4\alpha_a}\left[\frac{\dot{q}^a}{N}-\beta_aq^a-\gamma_a\right]^2
-NV(q^a)\right),
\ee
where the ``velocity'' $\dot{q}^a:=dq^a/d\tau'$ is the continuous limit of $\Delta q_n^a/\Delta\tau'_{n+1}$.

Therefore, in the special case that the Hamiltonian is a quadratic polynomial in $p_a$, the transition amplitude admits a path integral formalism over the configuration space, whereby the functional integration over $N$ is modified as $\int\mathcal{D}\sqrt{N}$. This is called the \emph{configuration space path integral} or \emph{Feynman's path integral}. The configuration space path integral \eqnref{eqn:Feynman's path integral} sums over all arbitrary paths $\gamma\in\mathcal{C}$ whose endpoints are fixed at $q'^a$ and $q^a$, and each path contributes with a phase, which is identical to the Lagrangian function as given in \eqnref{eqn:cl Lagrangian} (divided by $\hbar$). The functional variations on \eqnref{eqn:Feynman's path integral} with respect to $\sqrt{N}$ and $q^a$ yield the classical Hamiltonian constraint and equation of motion as in \eqnref{eqn:var to N on L} and \eqnref{eqn:var to q on L}.\footnote{Note that $\delta W/\delta\sqrt{N}=2\sqrt{N}\,\delta W/\delta N$.} This shows again that the stationary solution is the classical solution and thus \eqnref{eqn:W approx} is a good approximation.

\subsection{Timeless path integral with multiple constraints}\label{sec:with multiple constraints}
If there are multiple constraints and the constraint operators $\hat{H}^i$ commute, the projector is given by \eqnref{eqn:projector multiple constraints} and \eqnref{eqn:W q q'} can be directly generalized as\footnote{In this subsection, the repeated index $i$ is not summed unless $\sum_i$ is explicitly used.}
\be\label{eqn:W q q' with multiple H}
W(q^a,q'^a)
\sim \int_{-\infty}^\infty d\tau^1 \cdots \int_{-\infty}^\infty d\tau^k\
\opelem{q^a}{e^{-i\sum_{i=1}^k \tau^i \hat{H}^i}}{q'^a}.
\ee
If each $\hat{H}^i$ is a polynomial of $\hat{q}^a$ and $\hat{p}_a$ and Weyl ordered, the linear sum $\hat{H}'=\sum_i\tau^i\hat{H}^i$ is also a polynomial and Weyl ordered. Thus, by replacing $\tau$ with $1$ and $\hat{H}$ with $\hat{H}'$ in \eqnref{eqn:continuous path integral for kernel}, it can be shown
\ba
\opelem{q^a}{e^{-i\sum_i\tau^i \hat{H}^i}}{q'^a}&=&
\int\mathcal{D}q^a \int\mathcal{D}p_a\
\exp\left(\frac{i}{\hbar}\int_{\tilde{\gamma}}p_adq^a\right)\nn\\
&&\qquad\quad\times
\exp\left(-i \sum_i \int_{\tilde{\gamma}} \tau^iH^i(q^a(\bar{\tau}),p_a(\bar{\tau}))\,d\bar{\tau}\right),
\ea
where $\bar{\tau}$ is a parameter for the curve $\tilde{\gamma}$ with $q'^a=q^a(\bar{\tau}=0)$ and $q^a=q^a(\bar{\tau}=1)$. Redefining $\tau^i\Delta\bar{\tau}_n$ as $\Delta\tau^i_n$, we then have
\ba\label{eqn:kernel with multiple H}
\opelem{q^a}{e^{-i\sum_i\tau^i \hat{H}^i}}{q'^a}&=&
\int\mathcal{D}q^a \int\mathcal{D}p_a\
\exp\left(\frac{i}{\hbar}\int_{\tilde{\gamma}}p_adq^a\right)\nn\\
&&\qquad\quad\times
\exp\left(-i \sum_i \int_{\tilde{\gamma}} H^i(q^a(\tau^i),p_a(\tau^i))\,d\tau^i\right),
\ea

As in the case with a single constraint, the first exponential in \eqnref{eqn:kernel with multiple H} is independent of parametrizations for the curve $\tilde{\gamma}$, and for the second exponential we can play the same trick by summing over different parametrizations to get rid of the seemingly dependence on parametrizations. Following the same steps in \secref{sec:path integral general structure}, for each $i$, we have
\ba
&&\int d\tau^i \int \left[\mathcal{D}\Delta\tau^i\right]_{\sum\!\Delta\tau^i_n=\tau^i}
\opelem{q^a}{e^{i\tau^i \hat{H}^i}}{q'^a}\\
&=&
\int\mathcal{D}q^a \int\mathcal{D}p_a \int\mathcal{D}N_i\
\exp\left[\frac{i}{\hbar}
\int_{\tilde{\gamma}}\left(p_a\frac{dq^a}{d\tau'}
-N_i(\tau')H^i\right)d\tau'\right]
\ea
for a given arbitrary parametrization $\tau'$. After summed over $\left[\mathcal{D}\Delta\tau^i\right]_{\sum\!\Delta\tau^i_n=\tau^i}$ for each $i$, \eqnref{eqn:W q q' with multiple H} yields
\begin{subequations}\label{eqn:path integral with multiple H}
\ba
\label{eqn:path integral with multiple H 1}
W(q^a,q'^q)&\sim&
\int\mathcal{D}q^a \int\mathcal{D}p_a\, \prod_{i=1}^k\int\mathcal{D}N_i\,
\exp\left[\frac{i}{\hbar}
\int_{\tilde{\gamma}}\left(p_a\frac{dq^a}{d\tau'}
-\sum_{i=1}^k N_iH^i\right)d\tau'\right]\\
\label{eqn:path integral with multiple H 2}
&\sim&
\int\mathcal{D}q^a \int\mathcal{D}p_a\, \prod_{i=1}^k\delta[H^i]\,
\exp\left[\frac{i}{\hbar}
\int_{\tilde{\gamma}} p_a dq^a \right],
\ea
\end{subequations}
which is the direct generalization of \eqnref{eqn:path integral 3} and \eqnref{eqn:path integral 4}. In the path integral, each path in $\Sigma$ contributes with a phase, which is the classical action given in \eqnref{eqn:cl action with N} divided by $\hbar$. Functional variation on \eqnref{eqn:path integral with multiple H 1} with respect to $N_i$, $q^a$ and $p_a$ again yields the classical equations \eqnref{eqn:cl constraint} and \eqnref{eqn:Hamilton eqs}.

\section{Summary and discussion}\label{sec:discussion}
Starting from the canonical formulation in \cite{Rovelli:book}, the timeless path integral for relativistic quantum mechanics is rigorously derived. Given in \eqnref{eqn:path integral 4}, the transition amplitude is formulated as the path integral over all possible paths in the constraint surface $\Sigma$ (through the confinement by the delta functional $\delta[H]$), and each path contributes with a phase identical to the classical action $\int_{\tilde{\gamma}}p_adq^a$ divided by $\hbar$. The alternative expression is given in \eqnref{eqn:path integral 3}, which is the functional integral over all possible paths in the cotangent space $\Omega=T^*\mathcal{C}$ as well as over the Lagrange multiplier $N$. The timeless path integral manifests the timeless feature of relativistic quantum mechanics, as the parametrization for paths has no physical significance. For the special case that the Hamiltonian constraint $H(q^a,p_a)$ is a quadratic polynomial in $p_a$, the transition amplitude admits the timeless Feynman's path integral over the paths in the configuration space $\mathcal{C}$, as given in \eqnref{eqn:Feynman's path integral}.

The formulation of timeless path integral is intuitively appealing and advantageous in many respects as it generalizes the action principle of relativistic classical mechanics by replacing the classical notion of a single trajectory with a sum over all possible paths. It is easy to see that the classical solution contributes most to the transition amplitude and thus \eqnref{eqn:W approx} is a good approximation for generic cases since the stationary solution is identical to the classical one. Various approximation methods developed in (conventional) path integral approaches can be readily adapted to the timeless description. Furthermore, timeless path integral offers a new perspective to see how the conventional quantum mechanics emerges from relativistic quantum mechanics within a certain approximation (as discussed in \secref{sec:deparametrizable systems}) and may provide new insight into the problem of time. Specifically, for strictly deparametrizable systems, relativistic quantum mechanics and conventional quantum mechanics are different at the level of kinematics but identical at the level of dynamics; for non-strictly deparametrizable systems, on the other hand, they are different for both kinematics and dynamics.

The formulation of timeless path integral can be directly extended for the dynamical systems with multiple constraints as given in \eqnref{eqn:path integral with multiple H}, if the constraint operators $\hat{H}^i$ commute. For the case that $\hat{H}^i$ do not commute but form a closed Lie algebra, the projector is no loner given by \eqnref{eqn:projector multiple constraints} but we have to invoke \eqnref{eqn:group averaging} to obtain the physical state, which leads to
\be\label{eqn:W q q' with multiple noncommuting H}
W(q^a,q'^a)
\sim \int d\mu(\vec{\theta})\,
\opelem{q^a}{e^{-i\vec{\theta}\cdot\hat{\vec{H}}}}{q'^a},
\ee
where $\theta^i$ are coordinates of the Lie group $G$ generated by $\hat{H}^i$.
Starting from \eqnref{eqn:W q q' with multiple noncommuting H} and following the similar techniques used in this paper, one expects to obtain the timeless path integral, but the measure of the functional integral $\prod_{i=1}^k\int\mathcal{D}N_i$ appearing in \eqnref{eqn:path integral with multiple H 1} would have to be nontrivially modified, as $\theta^i$ play the same role of $\tau^i$ in \eqnref{eqn:W q q' with multiple H} but now the nontrivial Haar measure $d\mu$ is involved and the nontrivial topology of $G$ has to be taken into account. Consequently, one also has to deal with the issues concerning global restrictions on the Lagrange multipliers that ensure compatibility with the boundary conditions of the path integral (see \cite{Hartle:1984ut,Brown:1992bq,Gryb:2008rz}). The BRST methods used in Appendix of \cite{Brown:1992bq} to derive the path integral for Jacobi's action could be adopted to handle the gauge redundancy by the multiple constraints. By elucidating the procedure of group averaging in light of BRST techniques, we wish to explicitly formulate the timeless path integral with multiple constraints in the future.
For the case that $\hat{H}^i$ do not form a closed Lie algebra, it is not clear how to construct the quantum theory which is free of quantum anomalies even in the canonical formalism. The timeless path integral may instead provide a new conceptual framework to start with for constructing the quantum theory.

Throughout this paper, we have focused on simple mechanical systems, but not field theories. In Section 3.3 of \cite{Rovelli:book}, the canonical treatment of classical field theories which maintains clear meaning in a general-relativistic context is presented as a direct generalization of the timeless formulation for relativistic classical mechanics (see also \cite{Gotay:1997eg} and references therein), and the corresponding quantum field theory is formulated in Section 5.3 of \cite{Rovelli:book}. The timeless path integral for relativistic quantum mechanics derived in this paper should be extended for the quantum field theory described in \cite{Rovelli:book}. We leave it for the future research.

In SFMs, the transition amplitude between two spin networks (i.e.\ quantum states of gravitational fields) is given by the sum (with appropriate weights) over all spin foams whose boundary consists of the given spin networks. As spin foams are two-complexes with colored faces and edges, the formulation of sum-over-spin-foams amplitudes is completely \emph{combinatorial} and any time-slicing through spin foams is of no physical significance. The timeless path integral developed in this paper makes no reference whatsoever to time-slicing and thus bears a close resemblance to the formulation of SFMs,\footnote{In SFMs, the general expression of the transition amplitude between two spin networks $s$ and $s'$ takes the form:
\[
W(s,s')=\sum_{\sigma} w(\Gamma) \prod_f\mathrm{dim}(j_f) \prod_eA_e(j_f,i_e) \prod_vA_v(j_f,i_e),
\]
where a spin foam $\sigma=(\Gamma,j_f,i_e)$ is given by a two-complex $\Gamma$ with a half-integer $j_f$ associated with each face $f$ and an intertwiner $i_e$ associated with each edge $e$; the boundary of $\sigma$ is given by the spin networks $s$ and $s'$; and $A_e$ and $A_v$ are the amplitudes associated with each edge $e$ and each vertex $v$. Compared to the timeless path integral \eqnref{eqn:path integral 4}, the functional integral over all paths $\int\mathcal{D}q^a\mathcal{D}p_a$ is analogous to the discrete sum over all spin foams $\sum_\sigma$ (more elaborately, $q^a$ corresponds to $\Gamma$ and $p_a$ to the coloring $(j_f,i_e)$); the kinematic weight factor $\exp(i\int_{\tilde{\gamma}}p_adq^a/\hbar)$ for each path $\tilde{\gamma}=(q^a,p_a)$ is analogous to the weight factor $w(\Gamma) \prod_f\mathrm{dim}(j_f) \prod_eA_e(j_f,i_e)$ for each spin foam $\sigma=(\Gamma,j_f,i_e)$; and the Hamiltonian constraint functional $\delta[H]\equiv \lim_{N\rightarrow\infty}\prod_{n=0}^{N-1} \delta(H(\bar{q}^a_n,p_{an}))$ is analogous to the product of vertex amplitudes $\prod_vA_v(j_f,i_e)$, as the Hamiltonian of LQG acting on the nodes of spin networks gives rise to the vertices of spin foams at which edges branch. Note that the set $\{\delta(H(\bar{q}^a_n,p_{an}))\}$ indexed by $n$ is not ordered in any sense of time-slicing, and neither is the set $\{A_v(j_f,i_e)$\} indexed by $v$. The resemblance is marked except for the striking difference between the functional integral over continuous variables in \eqnref{eqn:path integral 4} and the discrete sum over discrete variables in SFMs.} as compared to other path integral treatments of timeless theories in which the notion of time-slicing does not fully disappear at the fundamental level (recall \footref{foot:clarification}). Therefore, as the timeless path integral in this paper is systematically derived from the well-defined canonical formulation of relativistic quantum mechanics, we expect it to provide new insight into the issues of the connection between LQG/LQC and SFMs. Extending the timeless path integral to field theories will make the resemblance to SFMs even stronger.

\begin{acknowledgements}
The author would like to thank Biao Huang for helpful discussions and anonymous reviewers of the previous manuscripts for bringing related works to his notice. This work was supported in part by the Grant No.\ 10675019 from the NSFC and the Grants No.\ 20080440017 and No.\ 200902062 from the China Postdoctoral Science Foundation. The major revision to the original manuscript owes much to the valuable comments elicited by Professor Wei-Tou Ni.
\end{acknowledgements}


\appendix

\section{Stationary phase approximation}\label{app:stationary approximation}
The timeless phase integral \eqnref{eqn:path integral 3} can be recast as
\be\label{eqn:path integral in action}
W(q^a,q'^a)\sim \int\mathcal{D}q^a \int\mathcal{D}p_a \int\mathcal{D}N\
\exp\left(\frac{i}{\hbar}S[\tilde{\gamma},N]\right),
\ee
where $S[\tilde{\gamma},N]\equiv \left.S[q^a,p_a,N]\right|_{\tilde{\gamma}}$ is the action given by \eqnref{eqn:cl action with N} for the path $\tilde{\gamma}$.
For the situations where the action for the classical solution $\tilde{\gamma}_\mathrm{cl}$ is much greater than $\hbar$, i.e.\ $S[\tilde{\gamma}_\mathrm{cl}]\gg\hbar$,\footnote{Note that the classical solution satisfies $H=0$ and thus $S[\tilde{\gamma}_\mathrm{cl}]$ is independent of the gauge choice of $N$.} the stationary phase approximation (\textit{\`{a} la} method of steepest descent) can be applied.
The main idea of the stationary phase method relies on the cancelation of the rapidly-varying phases (as $S[\tilde{\gamma}_\mathrm{cl}]/\hbar$ behaves as a huge number) of the paths $\tilde{\gamma}$ considerably deviated from the stationary solution, which is identical to the classical solution $\tilde{\gamma}_\mathrm{cl}$. Essentially, only the paths in the vicinity of the classical solution contribute in \eqnref{eqn:path integral in action}.

First, we functionally expand the action $S[\tilde{\gamma},N]$ in a Taylor series about the classical solution $\tilde{\gamma}_\mathrm{cl}$:
\ba\label{eqn:Taylor series}
&&S[\tilde{\gamma},N]\\
&=&
S[\tilde{\gamma}_\mathrm{cl}]
+\frac{1}{2!}\int d\tau\, d\tau' \left.\frac{\delta^2 S}{\delta\eta^A(\tau)\,\delta\eta^B(\tau')}\right|_{\tilde{\gamma}_\mathrm{cl}} \!\left(\eta^A(\tau)-\eta^A_\mathrm{cl}(\tau)\right)
\left(\eta^B(\tau')-\eta^B_\mathrm{cl}(\tau')\right)+\cdots,\nn
\ea
where $\eta^A=(q^a,p_b)$ and $\eta^A_\mathrm{cl}=(q^a_\mathrm{cl},p_{b\mathrm{cl}})$ is the classical solution. Note that the first-order functional derivatives vanish as the classical solution coincides with the stationary solution, and the second-order derivatives explicitly read as
\begin{subequations}\label{eqn:2nd derivatives}
\ba
\left.\frac{\delta^2S}{\delta q^a(\tau)\,\delta q^b(\tau')} \right|_{\tilde{\gamma}_\mathrm{cl}}
&=& -N(\tau) \left.\frac{\partial^2H}{\partial q^a(\tau)\,\partial q^b(\tau)}\right|_{\tilde{\gamma}_\mathrm{cl}} \delta(\tau-\tau'),\\
\left.\frac{\delta^2S}{\delta q^a(\tau)\,\delta p_b(\tau')} \right|_{\tilde{\gamma}_\mathrm{cl}}
&=& -N(\tau) \left.\frac{\partial^2H}{\partial q^a(\tau)\,\partial p_b(\tau)}\right|_{\tilde{\gamma}_\mathrm{cl}} \delta(\tau-\tau')
-\delta_a^b \frac{d}{d\tau}\delta(\tau-\tau'),\\
\left.\frac{\delta^2S}{\delta p_a(\tau)\,\delta p_b(\tau')} \right|_{\tilde{\gamma}_\mathrm{cl}}
&=& -N(\tau) \left.\frac{\partial^2H}{\partial p_a(\tau)\,\partial p_b(\tau)}\right|_{\tilde{\gamma}_\mathrm{cl}} \delta(\tau-\tau').
\ea
\end{subequations}
Taking \eqnref{eqn:2nd derivatives} into \eqnref{eqn:Taylor series} yields
\begin{subequations}\label{eqn:Taylor series 2}
\ba
S[\tilde{\gamma},N]
&\approx& S[\tilde{\gamma}_\mathrm{cl}]
+\int d\tau \frac{d(q^a-q^a_\mathrm{cl})}{d\tau}\,(p_a-p_{a\mathrm{cl}})\\
&&\quad
-\frac{1}{2}\int d\tau\, N(\tau) \left.\frac{\partial^2H}{\partial\eta^A(\tau)\,\partial\eta^B(\tau)}\right|_{\tilde{\gamma}_\mathrm{cl}} \!\left(\eta^A(\tau)-\eta^A_\mathrm{cl}(\tau)\right)
\left(\eta^B(\tau)-\eta^B_\mathrm{cl}(\tau)\right)\nn\\
&=:& S[\tilde{\gamma}_\mathrm{cl}]
+ \int d\tau \left(\delta p_a(\tau)\,\frac{d\,\delta q^a(\tau)}{d\tau} -N(\tau)\,\tilde{H}(\delta q^a,\delta p_a)\right),
\ea
\end{subequations}
where the perturbation variables $\delta\eta^A$ are defined as
\be
\delta\eta^A:=\eta^A-\eta^A_\mathrm{cl} \equiv (q^a-q^a_\mathrm{cl},p_b-p_{b\mathrm{cl}}) =:(\delta q^a,\delta p_b)
\ee
with $\delta q^a=0$ at the endpoints of $\delta\tilde{\gamma}:=\tilde{\gamma}-\tilde{\gamma}_\mathrm{cl}$,
and $\tilde{H}(\delta q^a,\delta p_a)$ is a function of $\delta q^a$ and $\delta p_a$ defined as
\be
\tilde{H}(\delta q^a,\delta p_a):=
\frac{1}{2}
\left.\frac{\partial^2H}{\partial\eta^A\,\partial\eta^B}\right|_{\tilde{\gamma}_\mathrm{cl}}
\! \delta\eta^A \delta\eta^B.
\ee

Next, substituting \eqnref{eqn:Taylor series 2} into \eqnref{eqn:path integral in action} and ignoring the higher order contributions, we obtain the stationary phase approximation:
\be\label{eqn:W approx single sol}
W(q^a,q'^a)\approx \zeta_{\tilde{\gamma}_\mathrm{cl}}\, e^{\frac{i}{\hbar}S[\tilde{\gamma}_\mathrm{cl}]}
\ee
where the weight factor $\zeta_{\tilde{\gamma}_\mathrm{cl}}$ is given by
\be\label{eqn:zeta}
\zeta_{\tilde{\gamma}_\mathrm{cl}}=
\int\mathcal{D}\delta q^a \int\mathcal{D}\delta p_a \int\mathcal{D}N\
\exp\left[\frac{i}{\hbar}\int_{\delta\tilde{\gamma}} d\tau \left(\delta p_a\frac{d\,\delta q^a}{d\tau} -N(\tau)\,\tilde{H}(\delta q^a,\delta p_a)\right)\right].
\ee
Note that \eqnref{eqn:zeta} takes the same form of \eqnref{eqn:path integral 3} except that $\eta^A$ is replaced by $\delta\eta^A$ and $H$ is replaced by $\tilde{H}$, implying that $\zeta_{\tilde{\gamma}_\mathrm{cl}}$ is independent of the parametrization of $\tau$ as it should be.
If there is only one single classical solution connecting $q'^a$ and $q^a$, the overall weight factor has no physical significance. On the other hand, if there are multiple classical solutions, \eqnref{eqn:W approx single sol} is generalized as \eqnref{eqn:W approx}, where the relative weights of $\zeta_{\tilde{\gamma}_i}$ are of physical importance and can be understood as the decoherence width of phases of the paths close to the classical solution $\tilde{\gamma}_i$.

Equations \eqnref{eqn:W approx single sol} and \eqnref{eqn:zeta} can be obtained less formally but more quickly by directly expanding the integrand in \eqnref{eqn:cl action with N}:
\begin{subequations}\label{eqn:expand integrand}
\ba
\label{eqn:expand integrand a}
&&p_a\frac{dq^a}{d\tau}-NH(q^a,p_a)\nn\\
&=&\left(p_{a\mathrm{cl}}+\delta p_a\right) \frac{d\left(q^a_\mathrm{cl}+\delta q^a\right)}{d\tau}
-NH(q^a_\mathrm{cl},p_{a\mathrm{cl}})
-N\left.\frac{\partial H}{\partial q^a}\right|_{\tilde{\gamma}_\mathrm{cl}}\!\delta q^a
-N\left.\frac{\partial H}{\partial p_a}\right|_{\tilde{\gamma}_\mathrm{cl}}\!\delta p_a\nn\\
&&\quad -N\frac{1}{2}
\left.\frac{\partial^2H}{\partial\eta^A\,\partial\eta^B}\right|_{\tilde{\gamma}_\mathrm{cl}}
\! \delta\eta^A \delta\eta^B +\cdots\\
\label{eqn:expand integrand b}
&=& \left(p_{a\mathrm{cl}}\frac{dq^a_\mathrm{cl}}{d\tau} -NH(q^a_\mathrm{cl},p_{a\mathrm{cl}})\right)
+\left(\delta p_a\frac{d\,\delta q^a}{d\tau} -N\tilde{H}(\delta q^a,\delta p_a)\right) +\cdots,
\ea
\end{subequations}
where from \eqnref{eqn:expand integrand a} to \eqnref{eqn:expand integrand b} we have applied \eqnref{eqn:Hamilton eqs} and integration by parts to eliminate the linear terms in $\delta\eta^A$. As $H(q^a_\mathrm{cl},p_{a\mathrm{cl}})=0$, taking \eqnref{eqn:expand integrand b} into \eqnref{eqn:path integral in action} immediately gives \eqnref{eqn:W approx single sol} with \eqnref{eqn:zeta}.\footnote{Carrying out $\int\mathcal{D}N$ in \eqnref{eqn:zeta} gives rise to a delta functional $\delta[\tilde{H}]$. However, it should be noted that $\tilde{H}(\delta\eta^A)\approx0$ is not the constraint for $\delta\eta^A$. Instead, given $H(\eta^A_\mathrm{cl})=0$, the Hamiltonian constraint $H(\eta^A)=0$ leads to $\left.\frac{\partial H}{\partial\eta^A}\right|_{\tilde{\gamma}_\mathrm{cl}}\!\delta\eta^A +\tilde{H}(\delta\eta^A)\approx0$ for $\delta\eta^A$. The linear terms in $\delta\eta^A$ are eliminated in \eqnref{eqn:expand integrand} and thus do not contribute to \eqnref{eqn:W approx single sol} and \eqnref{eqn:zeta}.}

\section{Path integral with time}\label{app:path integral}
In order to compare with the timeless path integral for relativistic quantum mechanics, in this appendix, we re-derive the path integral formalism (with time) for conventional quantum mechanics but include more generality: the (nonrelativistic) Hamiltonian $H_0$ is allowed to have explicit dependence on time, i.e., $\hat{H}_0=\hat{H}_0(\hat{q}^i,\hat{p}_i;t)$ with $t$ being the time parameter.
For a given Hamiltonian operator $\hat{H}_0$, the (nonrelativistic) transition amplitude is defined as
\be\label{eqn:def of G}
G(q^i,t;q'^i,t'):=\opelem{q^i}{\hat{U}(t-t')}{q'^i} =\opelem{q^i}{\mathcal{T}e^{-\frac{i}{\hbar}\int_{t'}^td\tau\hat{H}_0(\tau)}}{q'^i},
\ee
where $\hat{U}(t-t')$ is the evolution operator from $t'$ to $t$ and $\mathcal{T}$ represents \emph{time-ordering}.

\subsection{First case}\label{app:1st case}
We first consider the systems in which $[\hat{H}_0(\hat{q}^i,\hat{p}_i;t_1),\hat{H}_0(\hat{q}^i,\hat{p}_i;t_2)]=0$ for all $t_1,t_2$. For these systems, the time ordering $\mathcal{T}$ in \eqnref{eqn:def of G} is superfluous and thus
\be
G(q^i,t;q'^i,t')=\opelem{q^i}{e^{-\frac{i}{\hbar}\int_{t'}^td\tau\hat{H}_0(\tau)}}{q'^i}
\ee
For given $t'$ and $t$, let us introduce a parametrization sequence: $\tau_0=t',\, \tau_1,\,\tau_2,\cdots\!,\tau_{N-1},\,\tau_N=t$ with $\tau_n\in\mathbb{R}$, and define $\Delta\tau_n:=\tau_n-\tau_{n-1}$. The conditions on the endpoints ($\tau_0=t'$ and $\tau_N=t$) correspond to $\sum_{n=1}^N\Delta\tau_N=t-t'$.\footnote{In the literature, the parametrization sequence is normally chosen to be uniform, i.e.\ $\Delta\tau_n=(t-t')/N$. Here, we purposely keep it generic (non-uniform and even unordered) to be compared with the timeless path integral (see \footref{foot:parametrization}).}
For a given arbitrary small number $\epsilon$, by increasing $N$, we can always make the parameter sequence fine enough, i.e.\ $\mathrm{mesh}\{\tau_n\}\leq|\tau|/N\leq\epsilon$, such that $-i\int_{t'}^td\tau\hat{H}_0(\tau) =-i\lim_{N\rightarrow\infty}\sum_{n=0}^{N-1}\Delta\tau_{n+1}\hat{H}_0(\tau_n)$. Consequently, we can recast the transition amplitude as
\begin{subequations}\label{eqn:G}
\ba
&&G(q^i,t;q'^i,t')=\opelem{q^i}{e^{-\frac{i}{\hbar}\int_{t'}^td\tau\hat{H}_0(\tau)}}{q'^i}\nn\\
\label{eqn:G a}
&=&\opelem{q_N^i}{e^{-\frac{i}{\hbar}\Delta\tau_N\hat{H}_0(\tau_{N-1})}\, e^{-\frac{i}{\hbar}\Delta\tau_{N-1}\hat{H}_0(\tau_{N-2})} \cdots e^{-\frac{i}{\hbar}\Delta\tau_1\hat{H}_0(\tau_0)}}{q_0^i} +\mathcal{O}(\epsilon^2)\\
\label{eqn:G b}
&=&
\left(\prod_{n=1}^{N-1}\int d^{\,d-1}q_n^i\right)
\opelem{q_N^i}{e^{-\frac{i}{\hbar}\Delta\tau_N \hat{H}_0(\tau_{N-1})}}{q_{N-1}^i}
\opelem{q_{N-1}^i}{e^{-\frac{i}{\hbar}\Delta\tau_{N-1} \hat{H}_0(\tau_{N-2})}}{q_{N-2}^i}
\cdots\nn\\
&&\qquad\qquad\qquad\qquad\cdots
\opelem{q_1^i}{e^{-\frac{i}{\hbar}\Delta\tau_1 \hat{H}_0(\tau_0)}}{q_0^i} +\mathcal{O}(\epsilon^2),
\ea
\end{subequations}
where we have identified $q^i$=$q_N^i$ and $q'^i=q_0^i$ and in \eqnref{eqn:G b} inserted $N-1$ times the completeness relation
\be
\int d^{d-1}q^i \, \ket{q^i}\bra{q^i}
:= \int dq^1\cdots dq^{\,d-1} \, \ket{q^1,\cdots,q^{d-1}}\bra{q^1,\cdots,q^{d-1}}\,.
\ee
Disregarding difference of $\mathcal{O}(\epsilon^2)$, \eqnref{eqn:G} is formally identical to \eqnref{eqn:kernel} except that $\hat{H}$ is now replaced by $\hat{H}_0(\tau_n)/\hbar$ and $q_n^a$ replaced by $q_n^i$.
Therefore, taking the formal replacements:
\ba
&&d \rightarrow d-1, \qquad q_n^a \rightarrow q_n^i, \quad p_n^a \rightarrow p_n^i, \quad \bar{q}_n^a \rightarrow \bar{q}_n^i,\nn\\
&& \hat{H}(\hat{q}^a,\hat{p}_a) \rightarrow \hat{H}_0(\hat{q}^i,\hat{p}_i;t)/\hbar, \quad H(\bar{q}_n^a,p_{na}) \rightarrow H_0(\bar{q}_n^i,p_{ni};t)/\hbar,\nn\\
&&\tau'\rightarrow \tau, \quad \tau'=0\ \rightarrow\ \tau=t', \quad \tau'=\tau\ \rightarrow\ \tau=t,
\ea
and assuming that the Hamiltonian operator $\hat{H}_0(\hat{q}^i,\hat{p}_i;t)$ is a polynomial of $\hat{q}^i$ and $\hat{p}_i$ (while the parameter $t$ is viewed as coefficients) and is Weyl ordered, all the derivations from \eqnref{eqn:kernel} to \eqnref{eqn:continuous path integral for kernel} can be carried over in the obvious way. Therefore, parallel to \eqnref{eqn:discrete path integral for kernel} and \eqnref{eqn:continuous path integral for kernel}, the (nonrelativistic) transition amplitude is given by
\begin{subequations}\label{eqn:conventional path integral}
\ba
\label{eqn:conventional path integral a}
&&G(q^i,t;q'^i,t')\nn\\
&=&\lim_{N\rightarrow\infty}
\left(\prod_{n=1}^{N-1}\int d^{\,d-1}q_n^i\right)
\left(\prod_{n=0}^{N-1}\int\! \frac{d^{\,d-1}p_{ni}}{(2\pi\hbar)^{d-1}}\right)
\exp\left(\frac{i}{\hbar}\sum_{n=0}^{N-1}p_{ni}\Delta q_n^i\right)\nn\\
&&\quad\times
\exp\left(-\frac{i}{\hbar}\sum_{n=0}^{N-1}\Delta\tau_{n+1}H_0(\bar{q}_n^i,p_{ni};\tau_n)\right)\\
&\equiv&\int\mathcal{D}q^i \int\mathcal{D}p_i\
\exp\left(\frac{i}{\hbar}\int_{\tilde{\gamma}_0}p_idq^i\right)
\exp\left(-\frac{i}{\hbar}\int_{\tilde{\gamma}_0} H_0(q^i(\tau),p_i(\tau);\tau)\,d\tau\right),
\ea
\end{subequations}
where $\tilde{\gamma}_0$ are arbitrary paths in the cotangent space $\Omega_0=T^*\mathcal{C}_0$ which is given by $\tilde{\gamma}_0(\tau)=(q^i(\tau),p_i(\tau))$ and with the endpoints projected to the nonrelativistic configuration space $\mathcal{C}_0$ fixed by $q^i(\tau=t')=q'^i$ and $q^i(\tau=t)=q^i$.

It should be remarked that \eqnref{eqn:conventional path integral a} is given for a \emph{fixed} parametrization sequence $\tau_0=t',\, \tau_1,\,\tau_2,\cdots\!,\tau_{N-1},\,\tau_N=t$, which is not necessarily uniform or ordered. If we choose different parametrization sequences (fine enough), they nevertheless yield the same transition amplitude. No further sum over different parametrization sequences is necessary, as opposed to $\int\mathcal{D}\Delta\tau$ in \eqnref{eqn:path integral 1}, which results in $\int\mathcal{D}N$ in \eqnref{eqn:path integral 2}, for timeless path integral.
If the parametrization sequence is not an ordered partition of the interval $[t',t]$, the paths $\tilde{\gamma}_0(\tau)$ are envisioned turning back and forth in time within some periods. This still gives the same transition amplitude as that by an ordered partition, because in \eqnref{eqn:G a} any array segment turning back and forth in time gives rise to
\ba
&&e^{-\frac{i}{\hbar}\Delta\tau_{n_2}\hat{H}_0(\tau_{n_2})}\, e^{-\frac{i}{\hbar}\Delta\tau_{n_2-1}\hat{H}_0(\tau_{n_2-1})} \cdots e^{-\frac{i}{\hbar}\Delta\tau_{n_1}\hat{H}_0(\tau_{n_1})}\nn\\
&\rightarrow& e^{-\frac{i}{\hbar}\int_{\bar{\tau}}^{\bar{\tau}}\hat{H}_0(\tau)\,d\tau}+\mathcal{O}(\epsilon^2)
=1+\mathcal{O}(\epsilon^2),
\ea
whenever $\tau_{n_1}\approx\tau_{n_2}=\bar{\tau}+\mathcal{O}(\epsilon)$.

\subsection{Second case}\label{app:2nd case}
Secondly, let's consider the systems in which $[\hat{H}_0(\hat{q}^i,\hat{p}_i;t_1),\hat{H}_0(\hat{q}^i,\hat{p}_i;t_2)]\neq0$. For these systems, the time ordering $\mathcal{T}$ in \eqnref{eqn:def of G} is not superfluous but we have
\begin{subequations}
\ba
&&\mathcal{T}e^{-\frac{i}{\hbar}\int_{t'}^td\tau\hat{H}_0(\tau)}\nn\\
&:=&
1+\frac{(-i)}{\hbar}\int_{t'}^{t}dt_1\hat{H}_0(t_1) +\frac{(-i)^2}{2!\,\hbar^2}\int_{t'}^{t}dt_1dt_2\mathcal{T}\left\{\hat{H}_0(t_1)\hat{H}_0(t_2)\right\}
+\cdots\nn\\
&\equiv&
1+\frac{(-i)}{\hbar}\int_{t'}^{t}dt_1\hat{H}_0(t_1) +\frac{(-i)^2}{\hbar^2}\int_{t'}^{t}dt_1\int_{t'}^{t_1}dt_2\, \hat{H}_0(t_1)\hat{H}_0(t_2)\nn\\
&&\quad +\frac{(-i)^3}{\hbar^3}\int_{t'}^{t}dt_1\int_{t'}^{t_1}dt_2\int_{t'}^{t_2}dt_3\, \hat{H}_0(t_1)\hat{H}_0(t_2)\hat{H}_0(t_3)
+\cdots\\
\label{eqn:time-ordered exp}
&=&\lim_{N\rightarrow\infty}
\left(
1+\frac{(-i)}{\hbar}\sum_{n_1=0}^{N-1}\Delta\tau_{n_1+1}\hat{H}_0(\tau_{n_1}) +\frac{(-i)^2}{\hbar^2} \sum_{n_1=0}^{N-1}\Delta\tau_{n_1+1} \sum_{n_2=0}^{n_1-1}\Delta\tau_{n_2+1} \hat{H}_0(\tau_{n_1})\hat{H}_0(\tau_{n_2})
\right.\nn\\
&&
\left.
\qquad\qquad +\frac{(-i)^3}{\hbar^3} \sum_{n_1=0}^{N-1}\Delta\tau_{n_1+1} \sum_{n_2=0}^{n_1-1}\Delta\tau_{n_2+1} \sum_{n_3=0}^{n_2-1}\Delta\tau_{n_3+1} \hat{H}_0(\tau_{n_1})\hat{H}_0(\tau_{n_2})\hat{H}_0(\tau_{n_3})
+\cdots\right)\nn\\
&=&\lim_{N\rightarrow\infty}
\left(1-\frac{i}{\hbar}\Delta\tau_N\hat{H}_0(\tau_{N-1})\right) \left(1-\frac{i}{\hbar}\Delta\tau_{N-1}\hat{H}_0(\tau_{N-2})\right)\cdots \left(1-\frac{i}{\hbar}\Delta\tau_1\hat{H}_0(\tau_{0})\right),
\ea
\end{subequations}
where, in the discrete expression in \eqnref{eqn:time-ordered exp}, the parametrization sequence is ordered:
\be\label{eqn:ordered tau}
t'=\tau_0<\tau_1<\tau_2<\cdots<\tau_{N-1}<\tau_N=t.
\ee
Thus, by \eqnref{eqn:time-ordered exp}, we have
\be
\mathcal{T}e^{-\frac{i}{\hbar}\int_{t'}^t\hat{H}_0(\tau)d\tau}
=e^{-\frac{i}{\hbar}\Delta\tau_N\hat{H}_0(\tau_{N-1})}\, e^{-\frac{i}{\hbar}\Delta\tau_{N-1}\hat{H}_0(\tau_{N-2})} \cdots e^{-\frac{i}{\hbar}\Delta\tau_1\hat{H}_0(\tau_0)} +\mathcal{O}(\epsilon^2),
\ee
which is identical to the operator in the middle in \eqnref{eqn:G a} (up to $\mathcal{O}(\epsilon^2)$) except that the parametrization sequence $\{\tau_n\}$ is no longer generic but has to be ordered as in \eqnref{eqn:ordered tau}. Consequently, everything in the first case can be exactly repeated and thus we also obtain \eqnref{eqn:conventional path integral} for the second case, but this time $\{\tau_n\}$ is ordered and the paths $\tilde{\gamma}_0(\tau)$ are not allowed to move backward in time.


\begin{thebibliography}{99}


\bibitem{Rovelli:1989jn}
  C.~Rovelli,
  ``Time in quantum gravity: Physics beyond the Schr\"{o}dinger regime,''
  Phys.\ Rev.\  D {\bf 43}, 442 (1991).

\bibitem{Rovelli:1988qp}
  C.~Rovelli,
  ``Is there incompatibility between the ways time is treated in general relativity and in standard quantum mechanics?,''
  in {\it Proceedings, Conceptual Problems of Quantum Gravity}, (North Andover 1988), 126-140.

\bibitem{Rovelli:1990jm}
  C.~Rovelli,
  ``Quantum mechanics without time: A model,''
  Phys.\ Rev.\  D {\bf 42}, 2638 (1990).

\bibitem{Rovelli:1991ni}
  C.~Rovelli,
  ``Quantum evolving constants: Reply to comment on `Time in quantum gravity: An Hypothesis',''
  Phys.\ Rev.\  D {\bf 44}, 1339 (1991).

\bibitem{Reisenberger:2001pk}
  M.~Reisenberger and C.~Rovelli,
  ``Spacetime states and covariant quantum theory,''
  Phys.\ Rev.\  D {\bf 65}, 125016 (2002)
  [arXiv:gr-qc/0111016].

\bibitem{Marolf:2002ve}
  D.~Marolf and C.~Rovelli,
  ``Relativistic quantum measurement,''
  Phys.\ Rev.\  D {\bf 66}, 023510 (2002)
  [arXiv:gr-qc/0203056].

\bibitem{Hartle:1992as}
  J.~B.~Hartle,
  ``Space-time quantum mechanics and the quantum mechanics of space-time,''
  in \textit{Gravitation and Quantizations: Proceedings of the 1992 Les Houches Summer School},
  edited by B. Julia and J. Zinn-Justin, (North Holland, Amsterdam, 1995)
  [arXiv:gr-qc/9304006].


\bibitem{Isham:1992ms}
  C.~J.~Isham,
  ``Canonical quantum gravity and the problem of time,''
  arXiv:gr-qc/9210011.

\bibitem{Rovelli:book} C. Rovelli, \textit{Quantum Gravity}, (Cambridge University Press, Cambridge, 2004).



\bibitem{Hartle:1986yu}
  J.~B.~Hartle and K.~V.~Kuchar,
  ``Path integrals in parametrized theories: The Free relativistic particle,''
  Phys.\ Rev.\ D {\bf 34}, 2323 (1986).

\bibitem{Hartle:1984ut}
  J.~B.~Hartle and K.~V.~Kuchar,
  ``The role of time in path integral formulations of parametrized theories,''
  in \textit{Quantum Theory of Gravity: Essays in honor of the 60th birthday of Bryce S. DeWitt},
  edited by S. M. Christensen, (Adam Hilger, Bristol 1984), 315-326.

\bibitem{Brown:1989ne}
  J.~D.~Brown and J.~W.~York, Jr.,
  ``Jacobi's action and the recovery of time in general relativity,''
  Phys.\ Rev.\ D {\bf 40}, 3312 (1989).

\bibitem{Brown:1992bq}
  J.~D.~Brown and J.~W.~York, Jr.,
  ``Microcanonical functional integral for the gravitational field,''
  Phys.\ Rev.\ D {\bf 47}, 1420 (1993)
  [gr-qc/9209014].

\bibitem{Teitelboim:1989fi}
  C.~Teitelboim,
  ``Hamiltonian formulation of general relativity,''
  in \textit{Proceedings, Quantum cosmology and baby universes} (Jerusalem 1989), 1-63.

\bibitem{Gryb:2008rz}
  S.~B.~Gryb,
  ``Jacobi's principle and the disappearance of time,''
  Phys.\ Rev.\ D {\bf 81}, 044035 (2010)
  [arXiv:0804.2900 [gr-qc]].

\bibitem{Halliwell:1992nj}
  J.~J.~Halliwell and M.~E.~Ortiz,
  ``Sum over histories origin of the composition laws of relativistic quantum mechanics,''
  Phys.\ Rev.\ D {\bf 48}, 748 (1993)
  [gr-qc/9211004].

\bibitem{Halliwell:2009rw}
  J.~J.~Halliwell,
  ``Probabilities in quantum cosmological models: A decoherent histories analysis using a complex potential,''
  Phys.\ Rev.\ D {\bf 80}, 124032 (2009)
  [arXiv:0909.2597 [gr-qc]].

\bibitem{Halliwell:2012nj}
  J.~J.~Halliwell and J.~M.~Yearsley,
  ``Pitfalls of path integrals: Amplitudes for spacetime regions and the quantum Zeno fect,''
  Phys.\ Rev.\ D {\bf 86}, 024016 (2012)
  [arXiv:1205.3773 [gr-qc]].




\bibitem{Engle:2007uq}
  J.~Engle, R.~Pereira and C.~Rovelli,
  ``The loop-quantum-gravity vertex-amplitude,''
  Phys.\ Rev.\ Lett.\  {\bf 99}, 161301 (2007)
  [arXiv:0705.2388 [gr-qc]].

\bibitem{Freidel:2007py}
  L.~Freidel and K.~Krasnov,
  ``A new spin foam model for 4d gravity,''
  Class.\ Quant.\ Grav.\  {\bf 25}, 125018 (2008)
  [arXiv:0708.1595 [gr-qc]].

\bibitem{Engle:2007wy}
  J.~Engle, E.~Livine, R.~Pereira and C.~Rovelli,
  ``LQG vertex with finite Immirzi parameter,''
  Nucl.\ Phys.\  B {\bf 799}, 136 (2008)
  [arXiv:0711.0146 [gr-qc]].

\bibitem{Kaminski:2009fm}
  W.~Kaminski, M.~Kisielowski and J.~Lewandowski,
  ``Spin-foams for all loop quantum gravity,''
  Class.\ Quant.\ Grav.\  {\bf 27}, 095006 (2010)
  [arXiv:0909.0939 [gr-qc]].




\bibitem{Ashtekar:2009dn}
  A.~Ashtekar, M.~Campiglia and A.~Henderson,
  ``Loop quantum cosmology and spin foams,''
  Phys.\ Lett.\  B {\bf 681}, 347 (2009)
  [arXiv:0909.4221 [gr-qc]].

\bibitem{Ashtekar:2010ve}
  A.~Ashtekar, M.~Campiglia and A.~Henderson,
  ``Casting loop quantum cosmology in the spin foam paradigm,''
  Class.\ Quant.\ Grav.\  {\bf 27}, 135020 (2010)
  [arXiv:1001.5147 [gr-qc]].



\bibitem{Wipf:1993xg}
  A.~W.~Wipf,
  ``Hamilton's formalism for systems with constraints,''
  arXiv:hep-th/9312078.



\bibitem{Arthurs-Kelly}
  E.~Arthurs and J.~L.~Kelly,~Jr.,
  ``On the simultaneous measurement of a pair of conjugate observables,''
  Bell Sys. Tech. J. {\bf 44},  725 (1965).

\bibitem{Uffink}
  J.~Uffink,
  ``The joint measurement problem,''
  Int. J. Theor. Phys. {\bf 33}, 199-212 (1994).




\bibitem{Marolf:1995cn}
  D.~Marolf,
  ``Refined algebraic quantization: Systems with a single constraint,''
  arXiv:gr-qc/9508015.

\bibitem{Marolf:2000iq}
  D.~Marolf,
  ``Group averaging and refined algebraic quantization: Where are we now?,''
  arXiv:gr-qc/0011112.


\bibitem{Thiemann:1996ay}
  T.~Thiemann,
  ``Anomaly-free formulation of non-perturbative, four-dimensional  Lorentzian quantum gravity,''
  Phys.\ Lett.\  B {\bf 380}, 257 (1996)
  [arXiv:gr-qc/9606088].

\bibitem{Laudisa}
  F.~Laudisa,
  ``The EPR argument in a relational interpretation of quantum mechanics,''
  Foundations of Physics Letters, {\bf 14} (2) 119-132 (2001).


\bibitem{Wheeler}
  J.~A.~Wheeler,
  ``Frontiers of time,''
  in \textit{Problems in the Formulations of Physics}, edited by G. T. di Francia (North-Holland, Amsterdam, 1979).

\bibitem{Jacques}
  V.~Jacques \textit{et al.},
  ``Experimental realization of Wheeler's delayed-choice gedanken experiment,''
  Science {\bf 315} (5814), 966-968 (2007).

\bibitem{Dowker}
  F.~Dowker and A.~Kent, ``Properties of consistent histories,''
  Phys. Rev. Lett. {\bf 75}, 3038-3041 (1995).

\bibitem{Greiner:book}
  W.~Greiner and J.~Reinhardt, \textit{Field Quantization}, (Springer-Verlag, Berlin Heidelberg, 1996).

\bibitem{Simon:book}
  B.~Simon, \textit{Functional Integration and Quantum Physics}, (Academic Press, New York, 1979).

\bibitem{Gotay:1997eg}
  M.~J.~Gotay, J.~Isenberg and J.~E.~Marsden,
  ``Momentum maps and classical relativistic fields. I: Covariant field theory,''
  arXiv:physics/9801019.



\end{thebibliography}
\end{document}